\documentclass[11pt,a4paper]{article}
\pdfoutput=1
\usepackage{jheppub}

\usepackage{color}
\usepackage{amsmath}
\usepackage{pifont}   
\usepackage{verbatim}   
\usepackage{acronym} 
    
\usepackage{amsfonts}
\usepackage{amssymb}
\usepackage{mathrsfs} 
\usepackage{graphicx}
\usepackage{multirow} 
\usepackage{slashed} 
\usepackage{array}

\usepackage{graphicx}
\usepackage{epsfig}
\usepackage{lscape}
\usepackage{rotating}
\usepackage{epsf}
\usepackage{bm}
\usepackage{booktabs}
\usepackage{array}
\usepackage{caption}
\usepackage{subcaption}
\usepackage[utf8]{inputenc}
\usepackage{float}

\usepackage{multirow}
\usepackage{array,booktabs,colortbl,multirow}
\usepackage{colortbl,xcolor,colordvi,color}
\usepackage{rotating}
\allowdisplaybreaks

\newcommand{\beq}{\begin{align}}% can be used as {equation} or  {align}
\newcommand{\eeq}{\end{align}}
\def\be{\begin{equation}}
\def\ee{\end{equation}}
\def\bea{\begin{eqnarray}}
\def\eea{\end{eqnarray}}
\def\bitem{\begin{itemize}}
\def\eitem{\end{itemize}}
\newcommand{\bec}{\begin{center}}
\newcommand{\eec}{\end{center}}
\newcommand{\ba}{\begin{array}}
\newcommand{\ea}{\end{array}}
\newcommand{\cmrule}{\midrule[0.25mm]}

\title{Constraints on the relaxion mechanism with strongly interacting vector-fermions}

\affiliation{Instituto de F\'isica, Universidade de S\~ao Paulo, \\C.P. 66.318, 05315-970 S\~ao Paulo, Brazil}

\author{Hugues Beauchesne,}
\author{Enrico Bertuzzo, and}
\author{Giovanni Grilli di Cortona}

\emailAdd{hubea44@if.usp.br, bertuzzo@if.usp.br, ggrilli@if.usp.br}

\abstract{We study the experimental constraints on strongly interacting vector-fermions compatible with the relaxion mechanism and investigate any possible tuning. The focus is on a minimal model and low confinement scale. More precisely, we study bounds from electroweak precision tests, Higgs decay, Big Bang nucleosynthesis and direct collider searches. The effect of these new fermions on vacuum stability is also investigated. Combining our bounds, we show that the relaxion mechanism becomes increasingly constrained and fine-tuned as the confinement scale decreases. For example, a confinement scale of a few tens of MeVs would require tuning at the percent level.}

\begin{document}

\maketitle
\section{Introduction}

After the discovery of the Higgs boson, the question {\it How is the electroweak scale stabilized against radiative corrections?} has become more pressing than ever. Broadly speaking, the solutions to the problem can be divided into two categories: solutions advocating new {\it symmetries} (as in Supersymmetry~\cite{Golfand:1971iw, Volkov:1973ix, Wess:1974tw, Wess:1974jb, Ferrara:1974pu, Dimopoulos:1981zb}, Composite Higgs models~\cite{Bellazzini:2014yua,Panico:2015jxa} or models with Neutral Naturalness~\cite{Chacko:2005pe, Burdman:2006tz}) or solutions advocating new {\it dynamics} (as in relaxion models~\cite{Graham:2015cka, Espinosa:2015eda, Hardy:2015laa, Patil:2015oxa, Antipin:2015jia, Jaeckel:2015txa, Gupta:2015uea, Batell:2015fma, Matsedonskyi:2015xta, Marzola:2015dia, Choi:2015fiu, Kaplan:2015fuy, DiChiara:2015euo, Ibanez:2015fcv, Fonseca:2016eoo, Gertov:2016uzs, Fowlie:2016jlx, Evans:2016htp, Kobayashi:2016bue, Hook:2016mqo, Choi:2016luu, Flacke:2016szy, McAllister:2016vzi, Choi:2016kke, Evans:2017bjs, Kusenko:2014lra, Yang:2015ida, Kusenko:2014uta}). With no New Physics (NP) discovered at the LHC, conventional theories predicting new colored partners -such as Supersymmetry and Composite Higgs models- are becoming more and more constrained, with levels of tuning typically worse than $1\%$. 

On the other hand, the recently proposed relaxion mechanism appears to be quite promising. The stabilization of the Electroweak (EW) scale does not require colored particles, but rather a modified cosmological evolution in order to achieve the observed value of the Higgs mass parameter. The mechanism is easily sketched: a new scalar degree of freedom coupled to the Higgs boson (the relaxion) undergoes a slow roll evolution in the early universe, effectively {\it scanning} the Higgs squared mass parameter from the typical cut-off scale down to the EW Symmetry Breaking (EWSB) scale. Once the Higgs Vacuum Expectation Value (vev) is turned on, a vev-dependent back-reaction is triggered which stops the relaxion evolution, freezing the Higgs vev to the observed value. As shown in the original proposal~\cite{Graham:2015cka}, it is technically natural to stop the relaxion slow roll when the Higgs vev is close to the critical line separating the unbroken and broken phases of EW symmetry. Although conceptually simple, relaxion models are still in an early development stage, since no complete UV model has been presented so far. In this direction, supersymmetric models have been proposed to stabilize the EW scale all the way up to the Planck scale~\cite{Batell:2015fma,Evans:2016htp}, and clockwork models have been proposed to explain the transplankian field excursions needed to scan the Higgs squared mass parameter all the way down to the EW scale~\cite{Choi:2015fiu,Kaplan:2015fuy}. Despite these difficulties, it is interesting to analyse the consequences of the relaxion mechanism from an Effective Field Theory (EFT) point of view. In particular, a well motivated question is: how is the vev-dependent back-reaction generated?\footnote{A vev-dependent barrier is not the only possibility to stop the relaxion evolution. For instance, Ref.~\cite{Hook:2016mqo} uses particle production as friction source, while in Ref.~\cite{You:2017kah} the relaxion triggers the end of inflation which stops its evolution.} A simple possibility is to introduce new vector-like fermions coupled to the Higgs boson and charged under a new confining interaction. If the relaxion couples through an anomaly to the gauge bosons of the new gauge group, the required vev-dependent barrier is generated. In this setup, the theory can be natural for a cut-off scale as high as $\Lambda \simeq 10^8$ GeV if the fermion masses are around the EW scale and inflation is requested not to last so long as to reintroduce fine-tuning~\cite{Choi:2016luu}.\footnote{A similar configuration has been analyzed in Ref.~\cite{Arvanitaki:2016xds} in a framework in which the solution to the hierarchy problem is linked with a solution of the cosmological constant problem. In this case, however, the vector-like fermions do not form bound states.} 

In this context, the purpose of the present article is to study the experimental constraints on strongly interacting vector-fermions compatible with the relaxion mechanism and investigate any possible tuning. As it represents a minimal benchmark, we focus on the non-QCD model of Ref.~\cite{Graham:2015cka}. We analyze bounds coming from colliders (Electroweak Precision Tests (EWPT), Higgs decays and direct searches) and from cosmology (Big Bang Nucleosynthesis (BBN)). We also study the impact of this new strongly coupled sector on vacuum stability and show that some regions of parameter space require additional new physics below the previously estimated cutoff, though these regions will turn out to be ruled out by other constraints. The actual details of the physics however depend drastically on whether the new confinement scale is above or below the electroweak scale. The case of a confinement scale above the electroweak scale having already been studied in Refs.~\cite{Antipin:2015jia, Agugliaro:2016clv}, we concentrate on a lower confinement scale. This is in fact the more natural case, as will be further explained in Sec.~\ref{Sec:Model}. Combining the different experimental bounds, we find that the amount of parameter space available diminishes as the confinement scale is decreased and that the regions which are not already excluded require an increasing amount of fine-tuning for the relaxion mechanism to even work. For example, a confinement scale of a few tens of MeVs would require a tuning at the percent level.

This paper is organized as follow. We first review the non-QCD model of Ref.~\cite{Graham:2015cka} and discuss the decay channels of the different particles introduced. Experimental constraints are then presented. These include electroweak precision measurements, Higgs branching ratios, Big Bang Nucleosynthesis and direct collider searches. Several of these constraints will be similar to those of Ref. \cite{Joglekar:2012vc}. Possible vacuum stability issues are then studied. Finally, fine-tuning is discussed and all experimental bounds are combined together.

\section{Summary of the model}\label{Sec:Model}
We begin by summarizing the non-QCD model of Ref. \cite{Graham:2015cka} and highlighting some of its most salient features. The decay channels of the newly introduced particles are discussed afterward.

\subsection{Relaxation of the electroweak scale}\label{sSec:RelaxionExplanation}
The Standard Model is first extended by an axion $\phi$, which is also referred to as the relaxion, and a new strongly coupled gauge group $\mathcal{G}$. Neglecting parameters of order one, the evolution of the relaxion is governed by the Lagrangian
\begin{equation}\label{eq:LagrangianR}
  \mathcal{L}_{R}= -(-M^2+k\phi)H^\dagger H - V(k\phi) - \frac{1}{32\pi^2}\frac{\phi}{f_\phi}G^{\mu\nu}\tilde{G}_{\mu\nu},
\end{equation}
where $M$ is the cutoff of the theory, $H^T=(H^+,(h+iA^0)/\sqrt{2})$ the Higgs doublet, $G^{\mu\nu}$ the field strength of $\mathcal{G}$ and $\tilde{G}_{\mu\nu}$ its dual. The parameter $k$ is a spurion that parametrizes the breaking of the periodicity $\phi \to \phi + 2\pi f_\phi$ and is assumed to be small. The potential $V(k\phi)$, whose exact form is unimportant, leads to $\phi$ slowly rolling toward negative values.

The relaxion mechanism consists in having $\phi$ start at a large positive value of order $M^2/k$ and slowly roll down. This effectively scans over the Higgs mass. To provide a solution to the hierarchy problem, it is necessary for $\phi$ to stop rolling shortly after the Higgs mass square becomes negative and $h$ acquires an expectation value. This is done by introducing a back-reaction potential which only becomes relevant for non-zero expectation values of the Higgs.

In this non-QCD model, the back-reaction potential is generated by the introduction of new vector-like fermions charged under $\mathcal{G}$. The minimal content is two $SU(2)_L$ doublets $L$ and $L^c$ of weak hypercharge $-1/2$ and $+1/2$ respectively, and two singlets under all SM groups $N$ and $N^c$. The fields $L$ and $N$ are fundamentals under $\mathcal{G}$, while $L^c$ and $N^c$ are antifundamentals. The Lagrangian governing these new strongly interacting particles is
\begin{equation}\label{eq:LagrangianSC}
  \mathcal{L}_{SC}= -m_L L L^c - m_N N N^c - y H L N^c - \tilde{y} H^\dagger L^c N + \text{h.c.}
\end{equation}
As it corresponds to a particle charged under electroweak groups, $m_L$ must be above the electroweak scale while $m_N$ can be much lighter. When $h$ acquires an expectation value, the neutral component of $L$ mixes with $N$ and similarly for $L^c$ and $N^c$. This leads to two neutral Dirac fermions and a charged one. The lightest neutral fermion, which we label $n_1$, will be mostly composed of $N$ and its conjugate. Its mass $m_{n_1}$ will play a crucial role in the generation of the back-reaction potential. The heaviest one, which we refer to as $n_2$, will mostly consist of the neutral parts of the doublets and will be close in mass to the charged fermion labeled $C^-$. The mass of the latter is unaffected by the Yukawa interactions and thus remains $m_L$ at tree-level.

If some of the new fermions are light enough, condensation will take place at a scale $\Lambda=4\pi f$. The assumption we will be making throughout this article is that $m_L > \Lambda > m_{n_1}$, as we will justify shortly. Under this assumption, only $n_1$ forms a condensate. Using naive dimensional analysis \cite{Manohar:1983md}, this will generate a back-reaction term of
\begin{equation}\label{eq:BackReaction}
  V_{BR}\approx \Lambda_{BR}^4(h) \cos\left(\phi/f_\phi\right) \approx 4\pi f^3 \rho(h)\cos\left(\phi/f_\phi\right),
\end{equation}
where $\rho(h)$ corresponds at tree-level to $m_{n_1}$ and is given by
\begin{equation}\label{eq:rho}
  \rho(h) = c_0 + c_2 h^2 + \mathcal{O}(h^4),
\end{equation}
with $c_0 = m_N$ and $c_2 = y\tilde{y}/2m_L$ at leading order.\footnote{We assume, as will always be the case unless stated otherwise, that all parameters in (\ref{eq:LagrangianSC}) are real.} Once the mass square of the Higgs becomes negative, the amplitude of the back-reaction term increases with the Higgs vev until $\phi$ stops rolling. For this to be successful though, $c_0$ is required not to be too large compared to $c_2 v^2$, where $v\approx 246$~GeV is the current Higgs expectation value. Else, $\phi$ would stop rolling long before the Higgs reaches its correct expectation value. However, the parameter $c_0$ receives radiative corrections which need to be kept under control least a tuning be reintroduced. First, a correction of $\sim (y\tilde{y}/16\pi^2)m_L \ln(M/m_L)$ is generated from loops involving the doublet. Second, closing the Higgs loop in the second term of Eq.~(\ref{eq:rho}) contributes $\sim y\tilde{y}f^2/m_L$. Naturalness then requires
\begin{equation}\label{eq:ConditionsNaturalness}
  f \lesssim v \qquad \text{and} \qquad m_L \lesssim \frac{4\pi v}{\sqrt{\ln M/m_L}}.
\end{equation}
As collider constraints will force $m_L$ to be above several hundred GeV, this upper limit on $f$ severely restricts the range $m_L$ could take such that $n_2$ and $C^-$ condensate. This rather severe coincidence problem is why we choose to focus on $m_L > \Lambda$, though the results of Sec.~\ref{Sec:SummaryBounds} will show that the opposite relation is not entirely ruled out yet. The additional assumption $\Lambda > m_{n_1}$ simply comes from requiring $n_1$ to condensate so that the back-reaction term emerges and that the relaxion mechanism works. Taking these considerations and a few additional constraints into account,\footnote{These assumptions are dominance of classical rolling over de-Sitter quantum fluctuation, presence of enough Hubble friction to stop the relaxion once relative minimums start to form, dominance of the inflaton energy over the relaxion energy and reasonable length of inflation.} Ref. \cite{Choi:2016luu} estimates the cutoff to be
\begin{equation}\label{eq:Cutoff}
  M \lesssim 10^8 \, \text{GeV} \left(\frac{\Lambda_{BR}(v)}{10^3\text{GeV}}\right)^{4/5}\left(\frac{\mathcal{N}_e}{10^{26}}\right)^{1/10},
\end{equation}
where $\mathcal{N}_e$ is the number of e-folds during inflation which should be lower than $10^{26}$ to avoid severe fine-tuning problems in the inflation sector \cite{German:2001tz, Dine:2011ws, Iso:2015wsf}. Setting $m_{n_1}$ to the maximal value compatible with the relaxion mechanism and imposing Eq.~(\ref{eq:ConditionsNaturalness}), a cutoff as high as $10^8$~GeV can be reached. On the other hand, requesting the cutoff of the theory to be above a few TeV sets a lower limit on the confinement scale $\Lambda$ of $\mathcal{O}(10)$ MeV.

\subsection{Particle decays}\label{sSec:Decays}
The particles $n_2$ and $C^-$ will typically decay promptly if produced at colliders. In particular, the dominant decays are $C^- \to n_1 \, W^-$ and $n_2 \to n_1 \, Z(h)$. The decay $n_2 \to C^- \, W^+$, although possible, is phase-space suppressed and negligible for all sensible regions of parameter space. In all cases, the gauge bosons or the Higgs can potentially be forced to be off-shell.

Being strongly coupled and stable, $n_1$ will hadronize into so-called dark mesons and baryons. The baryons will remain stable, but the mesons will eventually decay back to Standard Model particles. We refer to the lightest meson made out of $n_1$ and its antiparticle as $\tilde{\eta}$. Its decay width can be estimated by first integrating out the doublets. In this situation, the symmetry breaking pattern is $U(1)_L\times U(1)_R \to U(1)_V$. Since the broken axial symmetry is anomalous, $\tilde{\eta}$ is not a pseudo-Goldstone boson and its mass $m_{\tilde{\eta}}$ should be around the confinement scale $\Lambda$. This analogy with the $\eta'$ meson of the Standard Model is the reason behind the name $\tilde{\eta}$.

Integrating out the doublets also generates two leading higher dimensional operator. The first one is the dimension five operator
\begin{equation}\label{eq:Op1}
  -\frac{y\tilde{y}}{m_L}H^\dagger H \overline{N}_D N_D,
\end{equation}
where $N_D$ is a neutral Dirac fermion with $N$ as its left-handed part and $N^c$ its right-handed part. This operator leads to $\tilde{\eta}$ decay via Higgs mixing. The second one is the dimension six operator
\begin{equation}\label{eq:Op2}
  -\frac{i}{2m_L^2}H^\dagger D_\mu H \overline{N}_D \gamma^\mu\left[(y^2+\tilde{y}^2) + (y^2-\tilde{y}^2)\gamma^5 \right] N_D + \text{h.c.}
\end{equation}
This operator contributes to $\tilde{\eta}$ decay via a virtual $Z$. Interpolating these two terms to the $\tilde{\eta}$ degree of freedom and using NDA leads to the following decay widths
\begin{equation}\label{eq:DecayWidthsZH}
  \begin{aligned}
    \Gamma_{\tilde{\eta}\to H^*\to\overline{f}f} &\sim \frac{y^2\tilde{y}^2 }{(4\pi)^3}\frac{m_f^2}{m_L^2m_h^4}m_{\tilde{\eta}}^5,\\
    \Gamma_{\tilde{\eta}\to Z^*\to\overline{f}f} &\sim \frac{(y^2-\tilde{y}^2)^2 g^2}{(4\pi)^3}\frac{m_f^2}{m_L^4}m_{\tilde{\eta}}^3,
  \end{aligned}
\end{equation}
where $m_f$ is the mass of some Standard Model fermions, $m_h$ is the mass of the Higgs boson, $g$ is the $SU(2)$ gauge coupling and a sum over fermions is implied. Which of these two decay channels dominates depends on the region of parameter space. Raising $m_L$ will increase the importance of the decay via Higgs mixing. However, decreasing $\Lambda$ will force the decay via Higgs mixing to go through lighter fermions which results in a Yukawa suppression.

An additional decay channel would be to two photons via the chiral anomaly or simply by loop diagrams. However, such decay would have to go through an operator of the form
\begin{equation}\label{eq:Op3}
  \frac{H^\dagger H F^{\mu\nu}F_{\mu\nu}\overline{N}_D N_D}{m_L^5},
\end{equation}
or with one of the field strengths replaced by its dual. Such operators would also have to be suppressed by some Yukawa couplings. A simple calculation shows that the decay width associated to this channel would be suppressed with respect to those of Eq.~(\ref{eq:DecayWidthsZH}) by an additional factor of $(m_{\tilde{\eta}}/m_L)^2$. The only way this decay channel could be important is if all decay channels to two fermions were kinematically forbidden. This is however highly unlikely, as this would require a confinement scale so low that the cutoff of the theory would be well below a TeV. The decay to two photons can therefore be neglected.

All in all, the decay width of $\tilde{\eta}$ can vary by many orders of magnitudes. Depending on the confinement scale $\Lambda$, $m_L$ and the new Yukawa couplings, $\tilde{\eta}$ can either decay spontaneously or be stable for macroscopic times. This will greatly affect the collider phenomenology and the cosmological bounds, as will be discussed in detail in the next section.

\section{Experimental constraints on the new fermions}\label{Sec:Constraints}
We now discuss the experimental bounds that the new fermion sector must satisfy. For simplicity, we assume the new strongly coupled group $\mathcal{G}$ to be $SU(N)$. Most results can however be easily translated to other groups. We also assume the confinement scale $\Lambda$ to be above $10$ MeV for reasons explained in Sec.~\ref{sSec:RelaxionExplanation}. Since the intent of this work is to study low confinement scales, we will generally limit ourselves to $\Lambda \lesssim 25 \, \mathrm{GeV}$, which corresponds to a cutoff scale $M \simeq 10^6$ GeV. For observables for which the confinement scale competes with the electroweak scale like EWPT or Higgs branching decay, neglecting the confinement scale then only introduces an error of at most a few percent.

\subsection{Electroweak precision tests}\label{sSec:EWPT}
Electroweak Precision Measurements (EWPM) play an important role in constraining physics beyond the Standard Model \cite{Amaldi:1987fu, Costa:1987qp, Langacker:1991an, Peskin:1991sw, Erler:1994fz, Altarelli:1990zd, Altarelli:1991fk, Grinstein:1991cd, Altarelli:1993sz, Barbieri:1999tm, Barbieri:2004qk}. In particular, the power of these indirect constraints has been shown by the agreement between the prediction of the top and the Higgs masses and their experimental values \cite{Erler:1994fz, ALEPH:2010aa, deBlas:2016ojx}. 

When new physics only modifies the vacuum polarization of gauge bosons (so-called oblique corrections), the impact of the new particles can be described in a first approximation by the three independent parameters $S$, $T$ and $U$ \cite{Peskin:1991sw}. However, if the new physics scale is close to or below the weak scale, one must also introduce the new parameters $V$, $W$ and $X$ \cite{Maksymyk:1993zm}. The six model independent oblique parameters are given by \cite{Maksymyk:1993zm}
\begin{align*}
\frac{\alpha(m_Z^2)}{4 s_W^2 c_W^2} S&= \frac{\Pi_{ZZ}(m_Z^2)-\Pi_{ZZ}(0)}{m_Z^2}-\frac{c_W^2-s_W^2}{s_W c_W}\Pi_{Z\gamma}'(0)-\Pi_{\gamma\gamma}'(0),\nonumber\\
\alpha(m_Z^2) T &= \frac{\Pi_{WW}(0)}{m_W^2}-\frac{\Pi_{ZZ}(0)}{m_Z^2},\nonumber\\
\frac{\alpha(m_Z^2)}{4 s_W^2 c_W^2} U&= \frac{\Pi_{WW}(m_W^2)-\Pi_{WW}(0)}{c_W^2 m_W^2}-\frac{\Pi_{ZZ}(m_Z^2)-\Pi_{ZZ}(0)}{m_Z^2}-\frac{2 s_W}{c_W} \Pi_{Z\gamma}'(0)-\frac{s_W^2}{c_W^2} \Pi_{\gamma\gamma}'(0),\nonumber\\
\alpha(m_Z^2) V &= 	\Pi_{ZZ}'(m_Z^2) - \frac{\Pi_{ZZ}(m_Z^2)-\Pi_{ZZ}(0)}{m_Z^2},\nonumber\\
\alpha(m_Z^2) W &= 	\Pi_{WW}'(m_W^2) - \frac{\Pi_{WW}(m_W^2)-\Pi_{WW}(0)}{m_W^2},\nonumber\\
\frac{\alpha(m_Z^2)}{s_W c_W} X &= 	\Pi_{Z\gamma}'(0) - \frac{\Pi_{Z\gamma}(m_Z^2)}{m_Z^2},
\end{align*}
where $\Pi_{XY}$ with $X,\,Y=\gamma,\,Z,\,W$ denotes the new physics contribution to the vacuum polarization amplitude of the gauge bosons, $\Pi_{XY}'(p^2)=d\Pi_{XY}(p^2)/dp^2$ and $s_W$ ($c_W$) are the tree-level SM values of the sine (cosine) of the weak mixing angle. 
Given the excellent agreement between the experimental measurements and the SM predictions, the oblique corrections cannot be larger than a few percent of the size of the leading order SM contribution. It is therefore a very good approximation to only include the leading order correction coming from new physics. An observable $A$ can thus be written as
\be
A = A_{SM} + A_{NP}(S, T, U, V, W, X),
\ee
where $A_{SM}$ is the SM prediction for $A$ (including radiative corrections) and $A_{NP}$ the leading correction coming from new physics.

In the model we are considering however, the new vector-like fermions not only contribute to the $\Pi_{XY}$ at loop-level, they also modify the $W$ and $Z$ boson widths already at tree-level. Indeed, when kinematically allowed, the $Z$ boson can decay to $\bar{n}_in_j$ or $C^-C^+$, while the W boson can decay to $C^\pm n_i$, where $i,j=1,2$. These tree-level corrections also affect the pole cross section for the process $e^+e^-\to Z\to$ hadrons, which is defined as
$$
\sigma_h^0=\frac{12 \pi}{m_Z^2}\frac{\Gamma_e \Gamma_\text{had}}{\Gamma_Z^2},
$$
where $\Gamma_e$ and $\Gamma_\text{had}$ are the partial widths of Z decay into $e^+e^-$ and hadrons.

\begin{table}[t]
{\footnotesize
\setlength\tabcolsep{5pt}
\begin{center}
\begin{tabular}{lcccc}
\toprule
& Ref. & Measurement & SM prediction \\
\cmrule
$m_W$ [GeV] & \cite{Group:2012gb} & $ 80.385 \pm 0.015 $ & $80.3610\pm0.0080$ \\ 
$\Gamma_{W}$ [GeV] & \cite{ALEPH:2010aa} & $ 2.085 \pm 0.042 $ &  $2.08849\pm0.00079$\\ 
$\sin^2\theta_{\rm eff}^{\rm lept}(Q_{\rm FB}^{\rm had})$\!\!\!\!\! & \cite{ALEPH:2005ab} & $ 0.2324 \pm 0.0012 $ & $0.23148\pm0.00012$\\
$P_{\tau}^{\rm pol}\!=\!\mathcal{A}_\ell$ & \cite{ALEPH:2005ab} & $ 0.1465 \pm 0.0033 $ & $0.14731\pm0.00093$ \\ 
$\Gamma_{Z}$ [GeV] & \cite{ALEPH:2005ab} & $ 2.4952 \pm 0.0023 $ & $2.49403\pm0.00073$ \\ 
$\sigma_{h}^{0}$ [nb] & \cite{ALEPH:2005ab} & $ 41.540 \pm 0.037 $ & $41.4910\pm0.0062$ \\
$R^{0}_{\ell}$ & \cite{ALEPH:2005ab} & $ 20.767 \pm 0.025 $ & $20.7478\pm0.0077$ \\ 
$A_{\rm FB}^{0, \ell}$ & \cite{ALEPH:2005ab} & $ 0.0171 \pm 0.0010 $ & $0.01627\pm0.00021$ \\ 
$\mathcal{A}_c$ & \cite{ALEPH:2005ab} & $ 0.670 \pm 0.027 $ & $0.66802\pm0.00041$ \\ 
$\mathcal{A}_b$ & \cite{ALEPH:2005ab} & $ 0.923 \pm 0.020 $ & $0.934643\pm0.000076$ \\ 
$A_{\rm FB}^{0, c}$ & \cite{ALEPH:2005ab} & $ 0.0707 \pm 0.0035 $ & $0.07381\pm0.00052$ \\ 
$A_{\rm FB}^{0, b}$ & \cite{ALEPH:2005ab} & $ 0.0992 \pm 0.0016 $ &  $0.10326\pm0.00067$\\ 
$R^{0}_{c}$ & \cite{ALEPH:2005ab} & $ 0.1721 \pm 0.0030 $ &  $0.172222\pm0.000026$ \\ 
$R^{0}_{b}$ & \cite{ALEPH:2005ab} & $ 0.21629 \pm 0.00066 $ & $0.215800\pm0.000030$ \\ 
\bottomrule
\end{tabular}
\end{center}
}
\caption{Experimental measurements and SM predictions for the set of EWPO considered.} 
\label{tab:EWPO}
\end{table}

\begin{figure}[t]
\begin{center}
    \includegraphics[width=0.57\textwidth, viewport = 0 0 324 328]{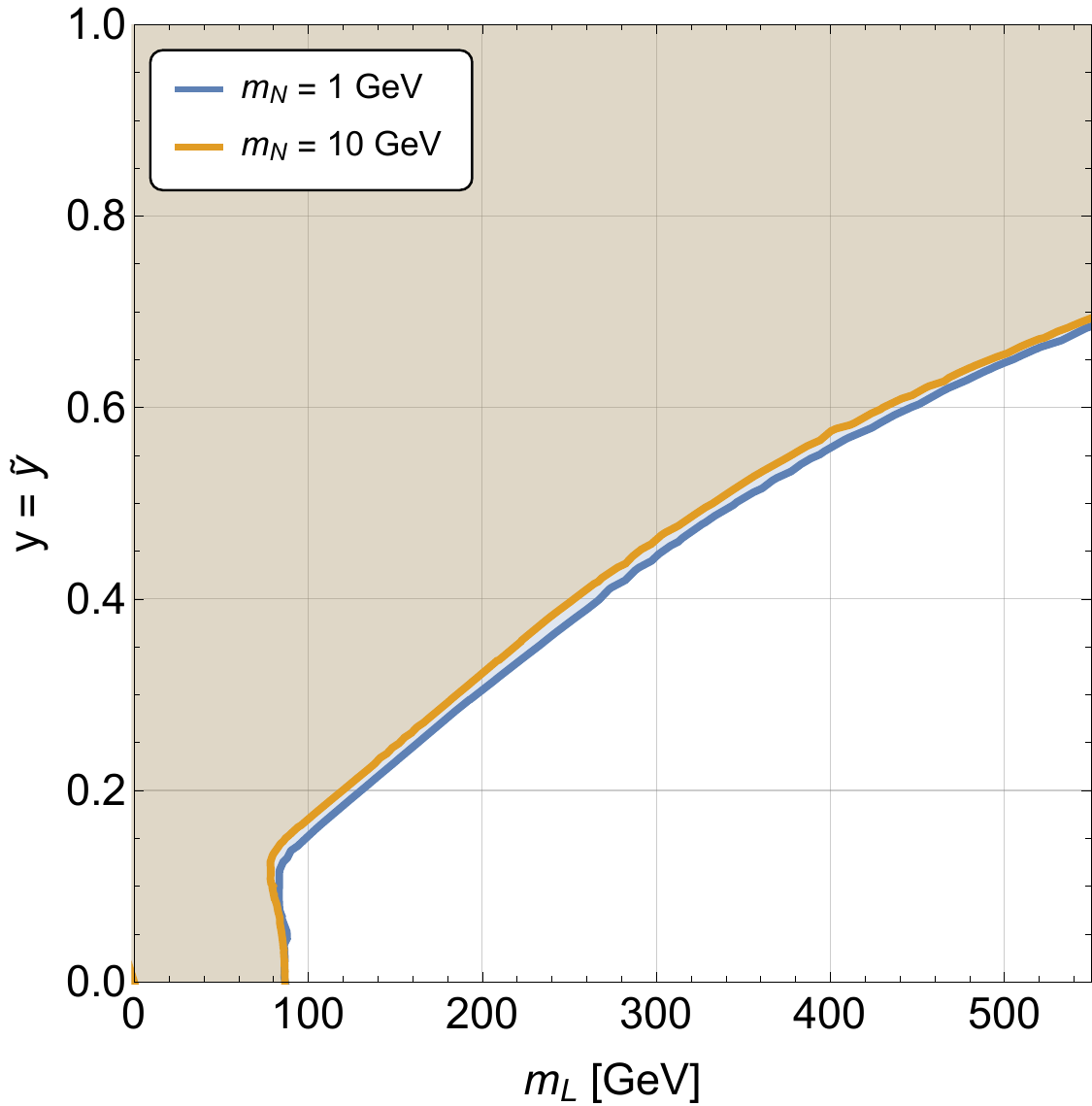}
  \end{center} 
\caption{Excluded region on the $(m_L,\,y=\tilde{y} )$ plane from electroweak precision observables, for different values of $m_N$ and for a new strong group $SU(3)$. The shaded region is excluded at  $95\%$ confidence level.}   \label{fig:EWPO}
\end{figure}

To compute how EWPM constrain the parameter space of the model, we performed a fit using the electroweak observables (EWPO) listed in Table~\ref{tab:EWPO}, where the third and fourth column show the experimental measurements and the SM predictions. We used the combined results from the SLD/LEP-I working groups for $\Gamma_Z$, $\sigma_h^0$, $P_\tau^{\mathrm{pol}}$, $\mathcal{A}_f$, $A_{FB}^{0,f}$ and $R_f^0$  \cite{ALEPH:2005ab} and from the LEP-II/Tevatron working groups for $\Gamma_W$ \cite{ALEPH:2010aa}. The most precise measurement of the $W$ boson mass was obtained by the Tevatron Electroweak Working Group \cite{Group:2012gb}. The SM predictions are taken from Table~2 of Ref. \cite{deBlas:2016ojx}, while the theoretical expressions for the observables can be derived from Refs. \cite{Maksymyk:1993zm,Burgess:1993mg,Burgess:1993vc,Bamert:1994yq}. We computed the $\chi^2$ function
\be
\chi^2(m_L,\,m_N,\,y\,,\tilde{y}) = \sum_{i=1}^{n_{\mathrm{obs}}} \left[\frac{A_i(m_L,\,m_N,\,y\,,\tilde{y})-\mu_i}{\delta \mu_i}\right]^2,
\ee
where $A_i$ are the observables listed in Table~\ref{tab:EWPO} and $\mu_i$ are their experimental values measured with an uncertainty $\delta \mu_i$. 

Fig.~\ref{fig:EWPO} shows the $95\%$ confidence level (CL) excluded region on the $(m_L,\,y=\tilde{y})$ plane from electroweak precision observables, taking a new $SU(3)$ strong group and for two different values of $m_N$, both much smaller than the electroweak scale. When the new Yukawa couplings $y=\tilde{y}$ go to zero, the constraints on $m_L$ become far less stringent. This is because the $S$ and $T$ parameters are related respectively to the dimension six operators $(H^\dagger \tau^I H)W_{\mu\nu}^I B^{\mu\nu}$ and $|H^\dagger D_\mu H|^2$. Since the coefficients of these operators could be calculated with the Higgses as external states, setting the Yukawa couplings to zero ({\it i.e.} no interactions between the new fermions and the Higgs doublet) results in these coefficients being null. The fact that not all constraints go away is because this argument does not apply to the higher dimensional operators that correspond to $V$, $W$ and $X$. In particular, for $y=\tilde{y} \lesssim 0.1$ the electroweak precision observables exclude vector-like fermions with $m_L\lesssim80$ GeV. As the new Yukawa couplings increase, the bound on $m_L$ gets stronger. For example, for $y=\tilde{y}\simeq 0.4$ and $m_N \lesssim 10$ GeV we get $m_L \gtrsim 260$ GeV.

\subsection{Higgs decays}\label{sSec:HiggsBR}
LHC studies of the Higgs production and decay provide additional constraints on the new fermions. Two effects must potentially be taken into account. First, the new fermions modify how the Higgs decays, mainly because of the presence of the new channels $h \to n_i \bar{n}_j$, where $i,\,j=1,\,2$. Since the collider constraints will show that $n_2$ needs to be considerably heavier than the Higgs, only decays involving $n_1$ are expected. These fermions then hadronize mostly into $\tilde{\eta}$'s. If these are short-lived, they lead to light jets that are not taken into account by Higgs studies (so-called {\it undetectable} objects \cite{Bechtle:2014ewa}) and their main effect is a reduction of the branching ratios to the most commonly studied channels. If they are long-lived, they additionally lead to Missing Transverse Energy (MET) that can be measured via recoil of the $Z$ boson in Higgs-strahlung processes \cite{Eboli:2000ze}. Decays to massive gauge bosons would also be affected at loop order, but we neglect this effect as it is subdominant. Second, the new fermions can change the cross section for Higgs production. However, this only affects vector boson fusion and is also a loop effect. We therefore also neglect the contributions of the new fermions to the cross section. 

A bound on Higgs decay to new physics that satisfies the requirements above can be read from Ref. \cite{Bechtle:2014ewa}. It is obtained from a global fit with \texttt{HiggsSignals} \cite{Bechtle:2013xfa, Bechtle:2008jh, Bechtle:2011sb, Bechtle:2013gu, Bechtle:2013wla} and is given by
\be
\mathrm{BR}(h\to\mathrm{NP})=\frac{\Gamma(h\to \mathrm{NP})}{\Gamma_h^{\mathrm{tot}}+\Gamma(h\to \mathrm{NP})}\leq20\%\quad\mathrm{at\,\,95\%\,\,CL},
\label{eq:BRh}
\ee
where 
$$
\Gamma(h\to \mathrm{NP})=\sum_{i,j}^{2}\Gamma(h\to n_i\bar{n}_j),
$$
and $\Gamma_h^{\mathrm{tot}}=4.1$ MeV is the theoretical prediction for the total decay width of the Higgs boson in the SM \cite{MelladoGarcia:2150771}. It is valid for both decay to undetectable objects or MET. Technically, the bound for decay to long-lived particles would be slightly stronger, but this effect is negligible on our results. We also verified using \texttt{HiggsSignals} that the inclusion of the most recent analyses do not change this result.

\begin{figure}[t]
\begin{center}
    \includegraphics[width=0.57\textwidth, viewport = 0 0 324 328]{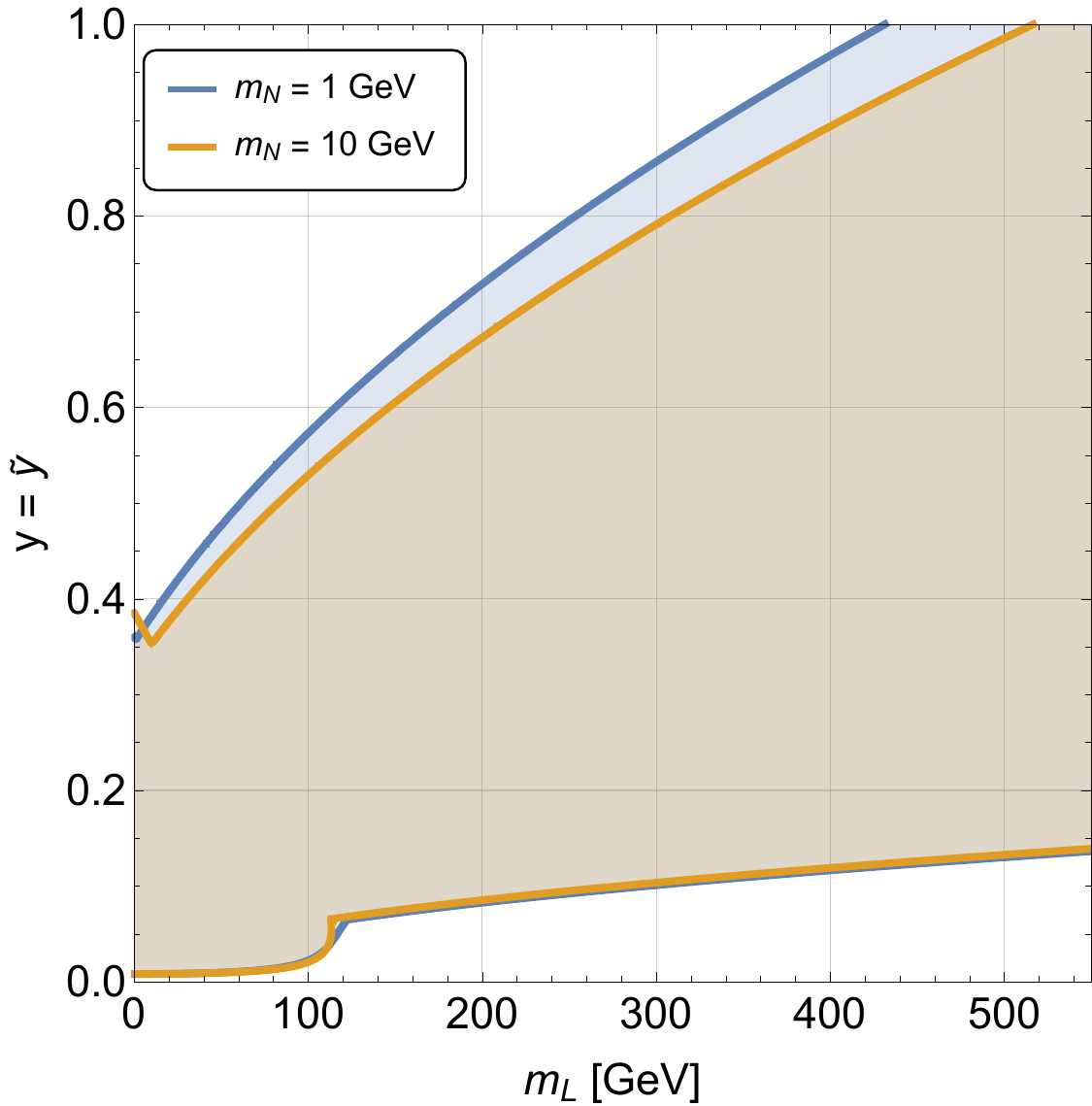}
  \end{center} 
\caption{Excluded region at $95\%$ confidence level from BR$(h\to n_i \overline{n}_j)<20\%$ in the $(m_L,\,y=\tilde{y})$ plane, for different values of $m_N$ and for a new strong group $SU(3)$.}   \label{fig:HiggsBR}
\end{figure}

We summarize in Fig.~\ref{fig:HiggsBR} the bounds from the Higgs to new physics branching ratio for $y=\tilde{y}$ for a new strong group $SU(3)$. It shows the constraints on $m_L$ and $y=\tilde{y}$ fixing $m_N$ to different values. For small new Yukawa couplings, the bound on $m_N$ and $m_L$ is slightly stronger than the one from electroweak precision observables. At very small new couplings $y,\,\tilde{y}\lesssim0.1$ there are no constraints for $m_L\gtrsim120$ GeV. In this region the only kinematically allowed decay is $h\to \bar{n}_1 n_1$, but the branching ratio is very suppressed due to the small couplings of $n_1$ to the Higgs boson. On the other hand, for very large values of the new couplings, the decay $h\to \mathrm{NP}$ is kinematically forbidden at tree-level, because $m_h < 2 m_{n_1}$. As a consequence there are no bounds from the tree-level Higgs branching ratio at large $y,\,\tilde{y}$. Loop corrections to the decay width to massive gauge bosons and the Higgs cross section could potentially set bounds for large enough $y=\tilde{y}$. However, we neglect this effect as this region is already excluded by electroweak precision observables (see Fig.~\ref{fig:EWPO}). For small $m_N$, which is the situation on which we are focusing, the bound is stronger than the one from electroweak precision observables for $y=\tilde{y}\lesssim 0.5$. However, as $m_N$ gets larger the region kinematically available for the decay $h\to \bar{n}_1n_1$ gets smaller and the bound fades away.

\subsection{Big Bang nucleosynthesis}\label{sSec:BBN}
As discussed in Sec.~\ref{sSec:Decays}, the $\tilde{\eta}$ mesons can easily be stable for macroscopic times. Being made out of fermions that are almost singlets, they will also decouple very early and therefore have a very large number density at decoupling. As such, they can potentially disturb Big Bang Nucleosynthesis, which would be observable from the abundance of light elements in the Universe. 

There are two ways in which BBN can be affected. If $\tilde{\eta}$ decays hadronically, it will inject relatively long-lived hadrons (e.g. $\pi^{\pm}$, $K^{0,\pm}$ and nucleons) which can modify the ratio of protons to neutrons, even for very small branching ratios to hadronic decay channels. For a branching ratio to hadronic channels of one percent or more, this leads to an upper bound on the decay time of $\tilde{\eta}$ of about $10^{-1}$ second for a $95\%$ confidence level (see for example \cite{Kawasaki:2004qu, Jedamzik:2006xz, Jedamzik:2009uy}). If $\tilde{\eta}$ instead decays radiatively (i.e. to photons or electrons), energetic photons can dissociate the nuclei of some light elements. The limit in this case is much less stringent, with an upper bound on the decay time of about $10^4$ seconds \cite{Cyburt:2002uv}.

\begin{figure}
  \begin{center}
   \includegraphics[width=0.57\textwidth, viewport = 0 0 450 436]{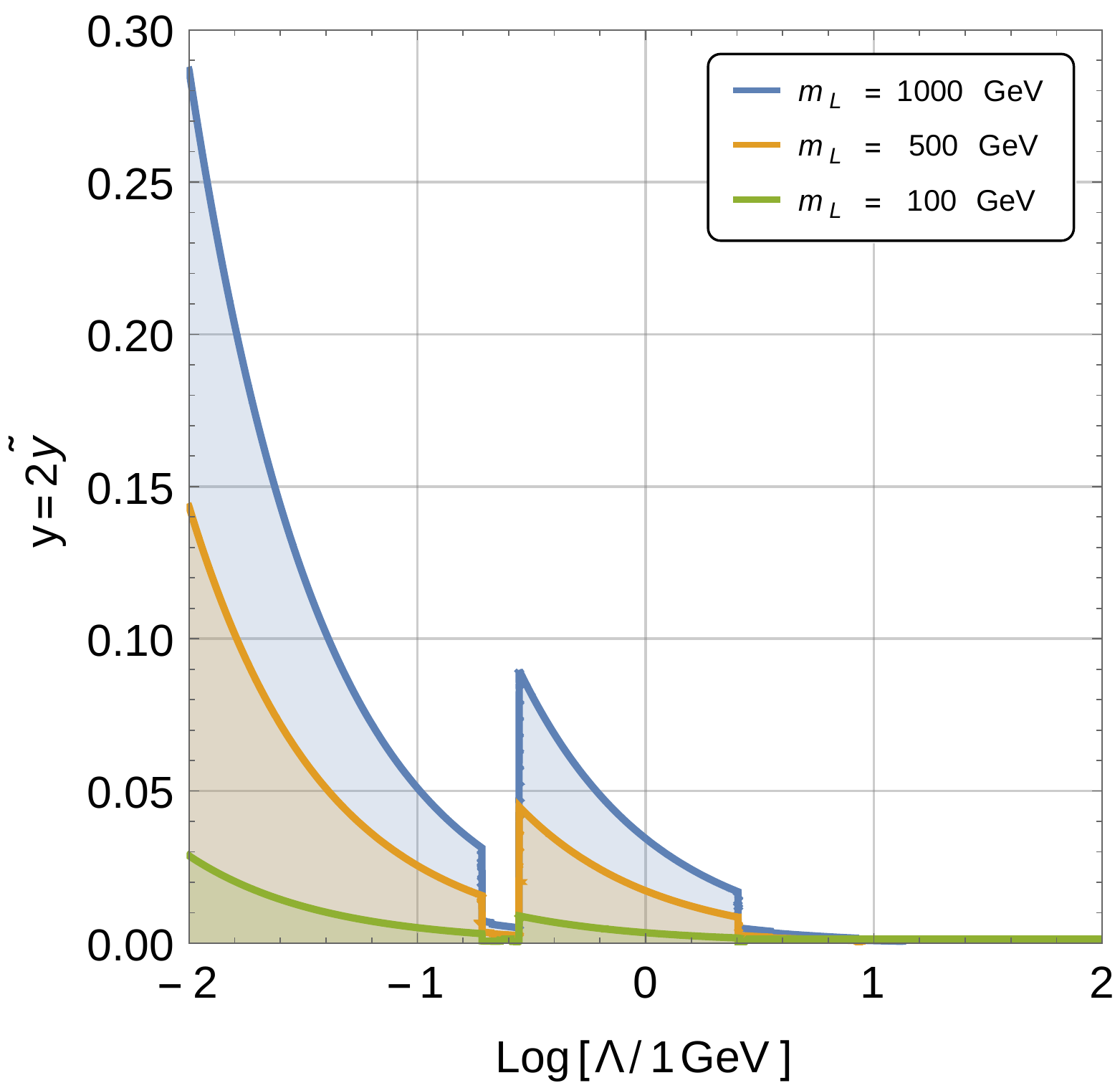}
   \caption{Limits from BBN on the Yukawa couplings and confinement scale for different values of $m_L$ and $N=3$. See text for the exclusion procedure.}
   \label{fig:BBN}
  \end{center}
\end{figure}

Fig.~\ref{fig:BBN} shows some constraints on the confinement scale and the new Yukawa couplings for different values of $m_L$. The criteria we use for judging whether a point of parameter space is excluded or not is as follows. If the confinement scale is above twice the mass of the charged pion, a point is excluded if the decay time is longer than $10^{-1}$ second. If it is lower than this, then a point is excluded if its decay time is longer than $10^4$ seconds. Of course, a more thorough treatment would lead to a smoother plot, but the behavior should not change much. This is also a conservative bound, as we neglect the decays $\tilde{\eta}\to \pi^+ e^- \bar{\nu}_e$ and its conjugate process which could still take place at a confinement scale between the masses of one and two charged pions albeit with a very suppressed rate. As is clear from Fig.~\ref{fig:BBN}, BBN provides a lower bound on the Yukawa couplings.

\subsection{Direct collider constraints}\label{sSec:Collider}
Collider data can be used to constrain a region of parameter space that is not covered by any of the previous bounds. This is the region where the Yukawa couplings are small enough to avoid the bounds of EWPT and Higgs branching ratios, but too large to disturb BBN. In a lot of ways, the collider phenomenology of the model is similar to that of electroweakinos, but with some small differences.

As long as they are light enough, any combination of a new fermion and a new antifermion can be produced via an $s$-channel photon, $W$ or $Z$. All decay chains will however end in $n_1$, which will then hadronize to dark baryons and mesons. What happens at this points depends on the decay length of $\tilde{\eta}$ and the phenomenology falls into three distinct categories. First, if $\tilde{\eta}$ decays promptly, $n_1$ will produce a jet that appears to originate from the principal vertex. Second, if $\tilde{\eta}$ decays inside the detector, an emerging jet will be produced \cite{Schwaller:2015gea}. Such objects would typically not be reconstructed by the different experiments. Finally, $\tilde{\eta}$ can decay outside the detector. The transverse momentum of $n_1$ is then simply converted to MET.

In this section, we will study each possible cases individually. These cases will then be combined together and with our previous constraints in Sec.~\ref{Sec:SummaryBounds}.

\subsubsection{Promptly decaying $\tilde{\eta}$}\label{ssSec:ColliderPrompt}
Promptly decaying $\tilde{\eta}$'s are characterized by all decay chains ending in a jet. In the region that is not already excluded by other constraints, $m_N$ must be at most of the order of a few GeV for the relaxion mechanism to work. Since $m_L$ will be forced to be considerably above the electroweak scale, this will lead to very boosted and narrow jets. The dark baryons will escape the detector leading to some MET (so-called {\it semi-invisible} jets \cite{Cohen:2015toa}). The energy carried by the baryons is usually of $\mathcal{O}(10\%)$ of the energy of the original $n_1$ for QCD like theories \cite{Carloni:2011kk} and diminishes as the rank of $SU(N)$ increases \cite{Witten:1979kh}. We will neglect this effect and the limits we obtain will therefore be slightly conservative.

\begin{figure}
  \begin{center}
   \includegraphics[width=0.38\textwidth, viewport = 0 0 285 240]{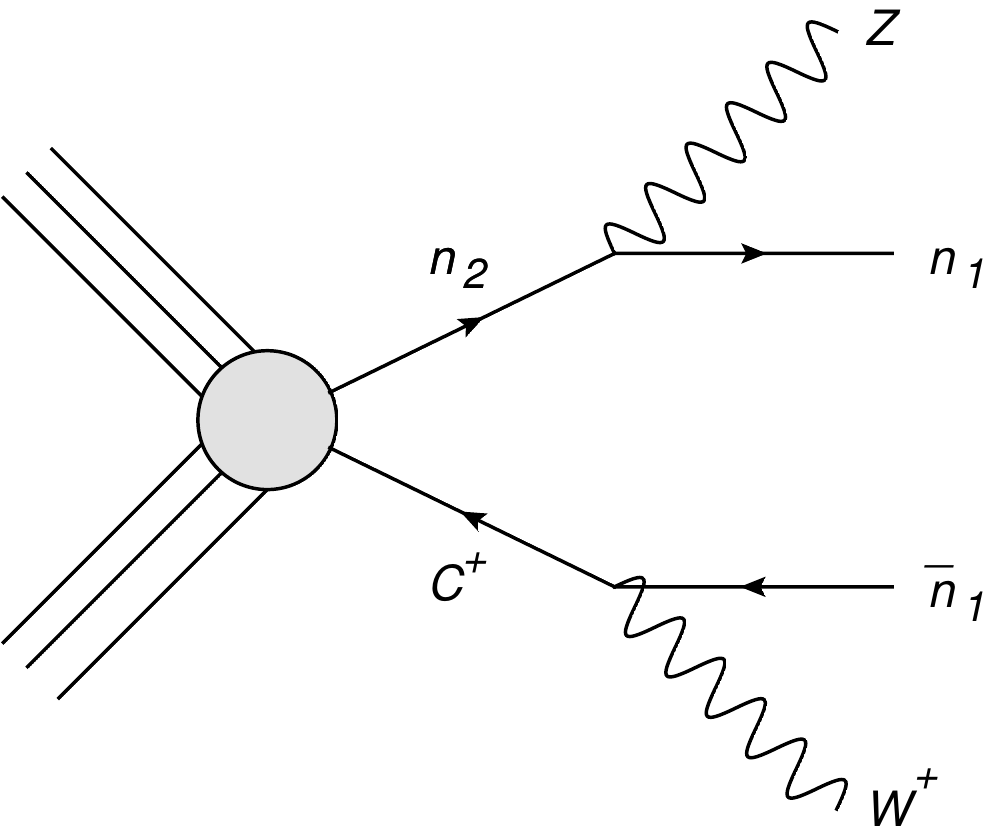}
   \caption{Topology contributing to the measurements of the $WZ$ differential cross section.}
   \label{fig:Feyn1}
  \end{center}
\end{figure}

\begin{figure}[t!]
  \centering
  \begin{subfigure}{0.47\textwidth}
    \centering
    \includegraphics[width=\textwidth, viewport = 0 0 450 366]{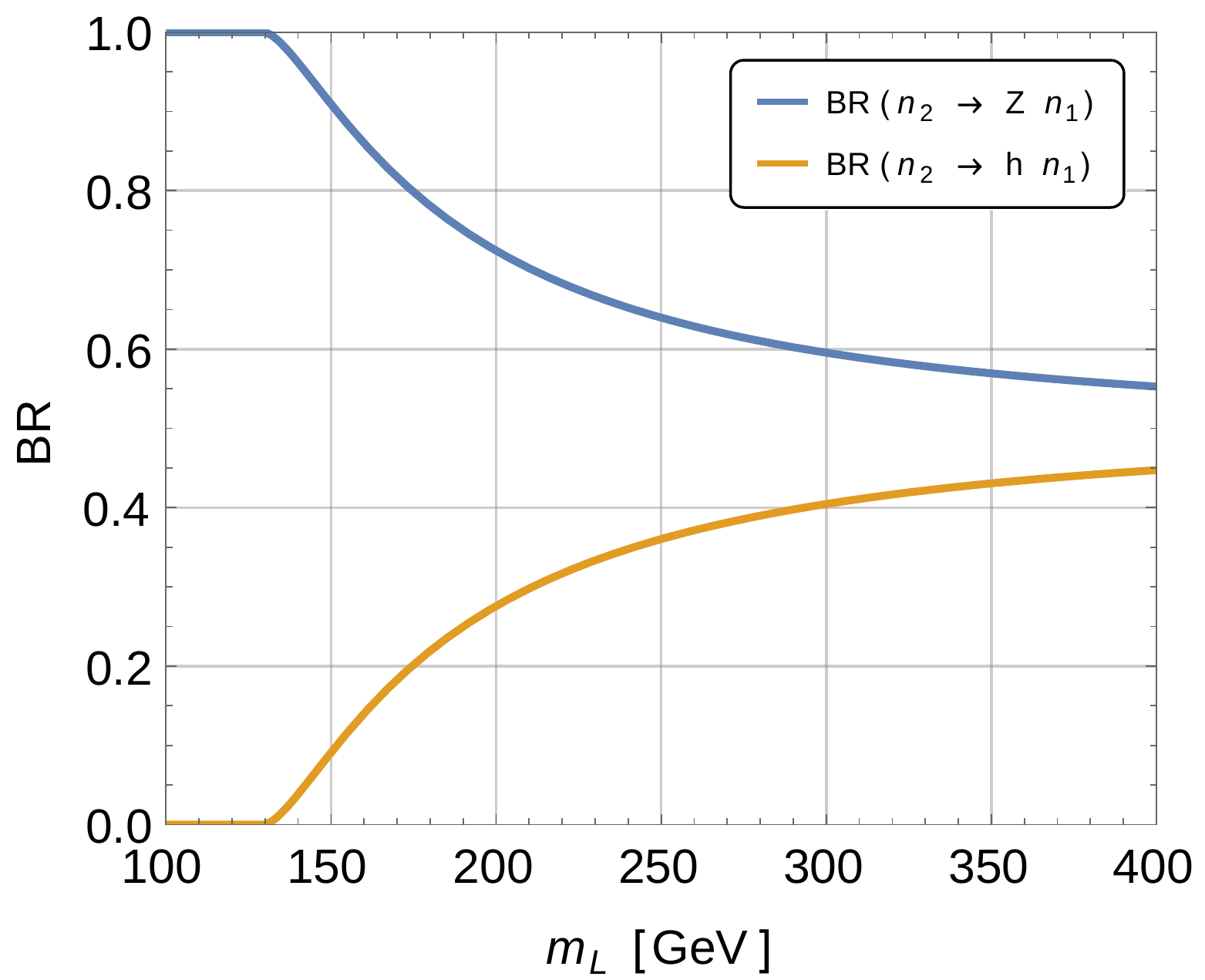}
    \caption{}
    \label{fig:BR}
  \end{subfigure}
  ~
    \begin{subfigure}{0.47\textwidth}
    \centering
    \includegraphics[width=\textwidth, viewport = 0 0 450 355]{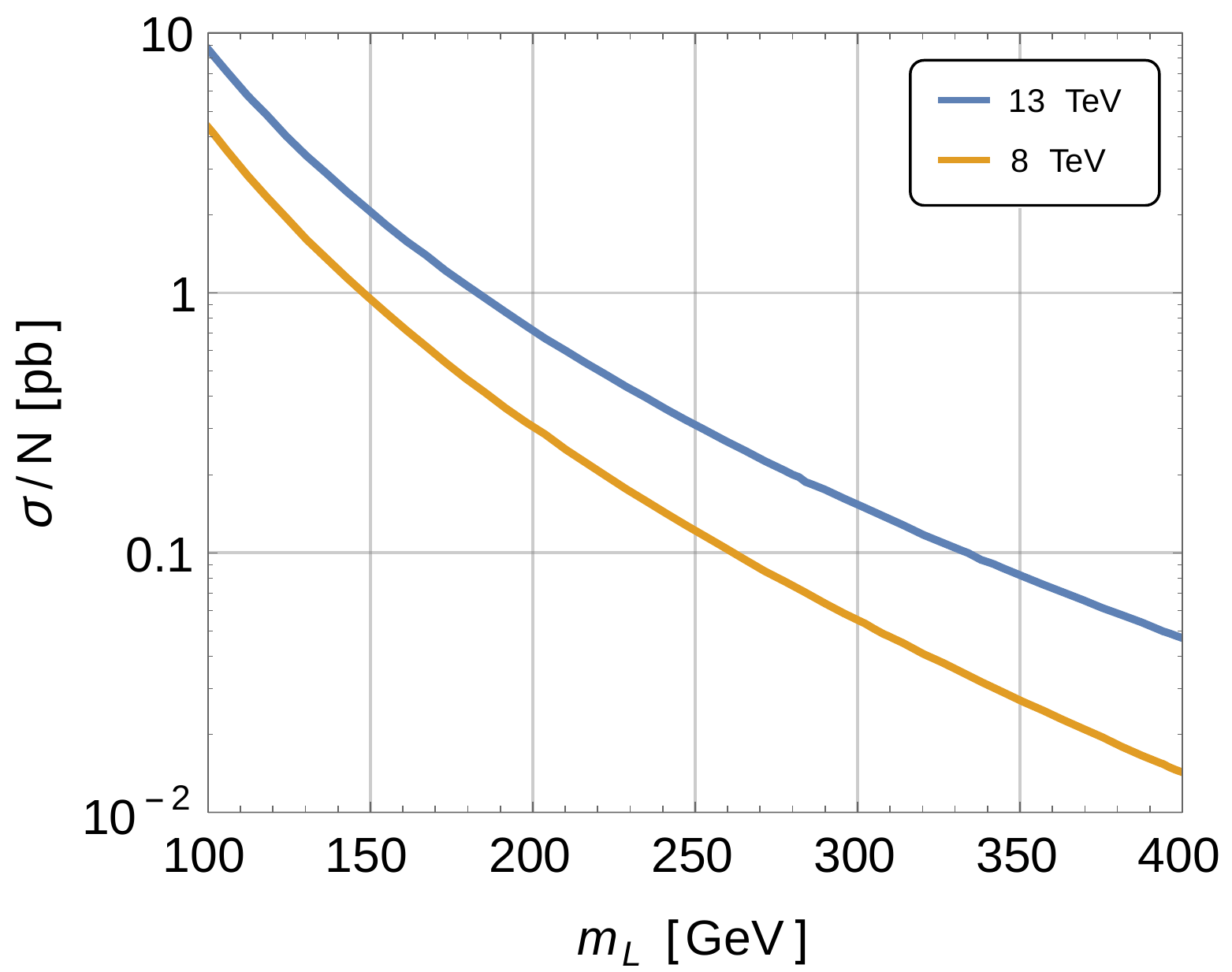}
    \caption{}
    \label{fig:CS}
  \end{subfigure}
\caption{(a) Branching ratios of $n_2$. (b) Tree-level cross sections $\sigma/N$ for $pp \to n_2C^+$ and its conjugate process at 8 and 13 TeV for $\mathcal{G}=SU(N)$. In both cases, the mass $m_N$ is set to 5 GeV and the Yukawa couplings to $y=\tilde{y}=0.1$. The confinement scale is assumed small.}\label{fig:BRCS}
\end{figure}

The production mechanism that we choose to focus on is the creation of $n_2C^+$ and the conjugate process. Of all the possible pairs of particles close to $m_L$ that can be produced, this process is typically the one with the largest cross section. As discussed in Sec.~\ref{sSec:Decays}, $C^+$ will decay to $\overline{n}_1$ and $W^+$. The neutral particle $n_2$ will decay to $n_1$ and either a $Z$ boson or a Higgs. We choose to concentrate on the decay to $Z$. This is simply because $n_2$ decays more often to $Z$ in the region of parameter space where collider bounds are important and that a $Z$ decay gives a signal that is cleaner and easier to constrain using experimental data. The signature of these events is then simply a $Z$, a $W$ and two jets, with topology shown in Fig.~\ref{fig:Feyn1}. An example of the branching ratios of $n_2$ is shown in Fig.~\ref{fig:BR}.

The pair production of $C^+C^-$ could also be used to constrain the model. However, the signature would now be two $W$'s and two jets, which would be difficult to distinguish from $t\overline{t}$. Alternatively, limits could be obtained from the pair production of $n_2\overline{n}_2$. Unfortunately, such limits would be suppressed by the fact that $Z$ decays less often to leptons than $W$ and by an additional branching ratio of $n_2$ to $Z$ and $n_1$. There is also less experimental measurements available in this case.

The search strategy that we adopt is to recast measurements of the differential cross section for $WZ$ production. Indeed, these measurements look for two Opposite Sign Same Flavor (OSSF) leptons that reconstruct to a $Z$, an additional lepton and a small amount of MET. In addition, several of these searches also bin their results in terms of the number of reconstructed jets. It is then possible to obtain bounds on the number of events with a leptonically decaying $W$ and $Z$ that also contain at least two jets, which is precisely our signature.

The measurements of the differential $WZ$ cross section that we focus on are Ref. \cite{Aaboud:2016yus} and Ref. \cite{Khachatryan:2016tgp}, both at 13 TeV. The first one is by ATLAS and corresponds to an integrated luminosity of 3.2~fb$^{-1}$. The second one is by CMS and corresponds to an integrated luminosity of 2.3~fb$^{-1}$. The searches of Refs. \cite{Aad:2016ett} and \cite{Khachatryan:2016poo} at 8 TeV could potentially have been used, but some crucial numbers about the differential cross sections were not made available to us. 

We wrote a series of codes to implement the cuts of Refs. \cite{Aaboud:2016yus, Khachatryan:2016tgp} and verified that we could reproduce their $WZ$ background estimates with good accuracy. Using an implementation of the model in Feynrules \cite{Alloul:2013bka}, we use \texttt{MadGraph5\_aMC@NLO} \cite{Alwall:2014hca} to generate 10000 events with the topology of Fig.~\ref{fig:Feyn1} and the gauge bosons decaying to light leptons. The events are then passed to \texttt{PYTHIA}~6 for hadronization \cite{Sjostrand:2006za}. Accurate simulation of hadronization in the dark sector is not an easy task. Since jets will be very narrow, we simply force \texttt{PYTHIA} to decay $n_1$ to two light quarks. This then almost always leads to a single jet for each $n_1$, which is the behavior that would be obtained from a more refined treatment. No information on jet substructure is ever used in the analysis. Detector simulation is handled with \texttt{Delphes} \cite{deFavereau:2013fsa} which includes FastJet \cite{Cacciari:2011ma}. Efficiencies for the reconstruction of electrons with different selection criteria (loose, medium, tight...) are taken from Refs.~\cite{ATLAS:2016iqc} and \cite{CMS:eff} for ATLAS and CMS respectively. Muon efficiencies, which are for all intents and purposes 100$\%$, are taken from Ref. \cite{Aad:2016jkr} in both cases. The cross section is calculated at tree-level using \texttt{MadGraph}. Examples at 8 and 13 TeV are shown in Fig.~\ref{fig:CS}. A $K$-factor of 1.2 is assumed when calculating experimental bounds. This is the typical $K$-factor associated to Higgsino pair production, which is almost identical in the limit of small Yukawa couplings \cite{Beenakker:1996ed}. All events are required to contain at least two jets.

\begin{figure}[t!]
  \centering
  \begin{subfigure}{0.47\textwidth}
    \centering
    \includegraphics[width=\textwidth, viewport = 0 0 450 366]{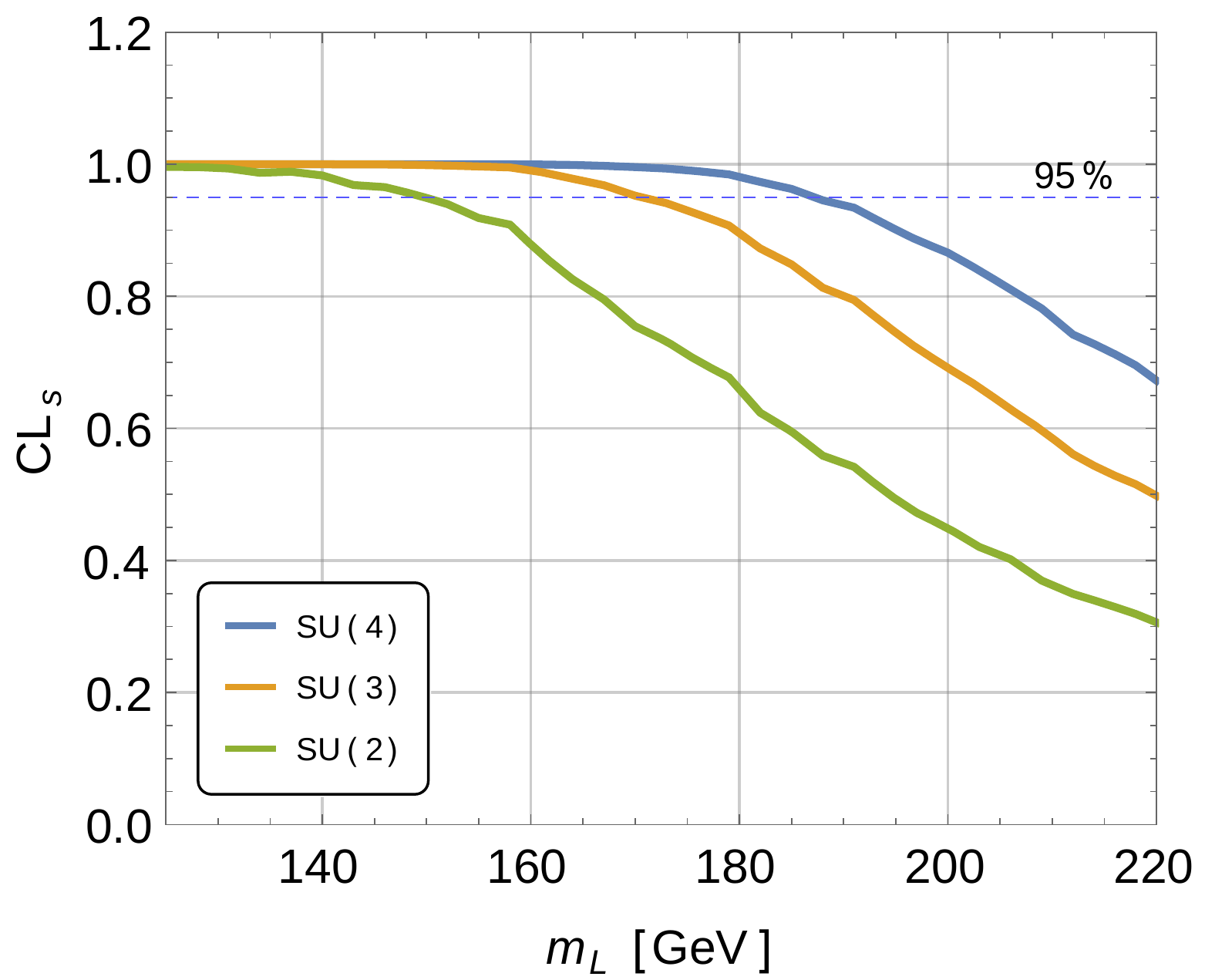}
    \caption{}
    \label{fig:CLsPrompt}
  \end{subfigure}
  ~
    \begin{subfigure}{0.47\textwidth}
    \centering
    \includegraphics[width=\textwidth, viewport = 0 0 450 366]{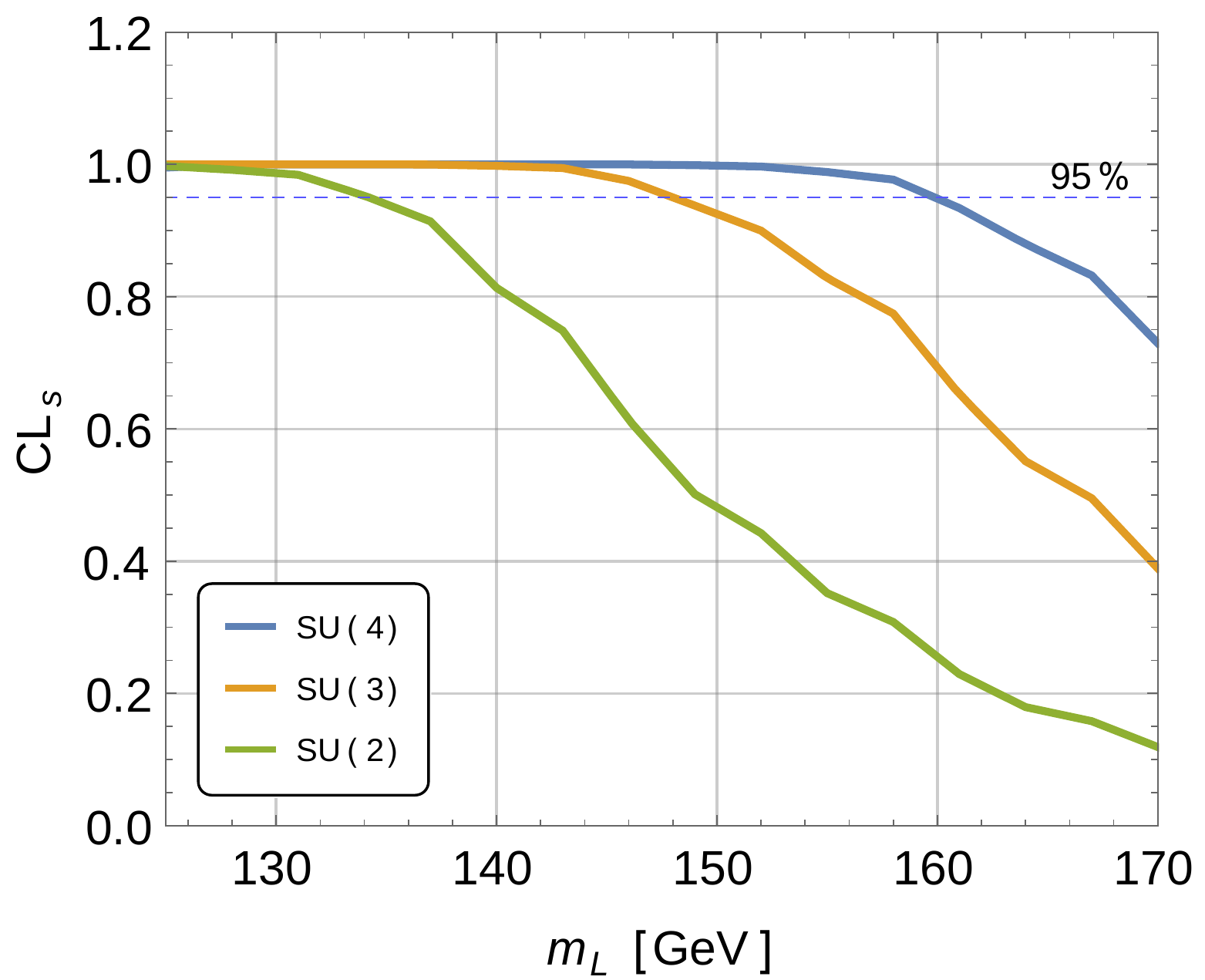}
    \caption{}
    \label{fig:CLsEmerging}
  \end{subfigure}
\caption{(a) $CL_s$ for promptly decaying $n_1$ as a function of $m_L$ and for different $SU(N)$. (b) $CL_s$ for $n_1$ decaying inside the detector as a function of $m_L$ and for different $SU(N)$. In both cases, the mass $m_N$ is set to 5 GeV and the Yukawa couplings to $y=\tilde{y}=0.1$. The confinement scale is assumed small.}\label{fig:CLs1}
\end{figure}

Considering the relevant searches correspond to very little integrated luminosity, we obtain limits by statistically combining them together using $CL_s$ techniques \cite{Read:2002hq, Junk:1999kv}. Since these searches represent different sets of data, the correlation should be negligible. Fig.~\ref{fig:CLsPrompt} shows the $CL_s$ as a function of $m_L$ for different $\mathcal{G}$ and for small Yukawa couplings and $m_N$. In the region of parameter space that is not already excluded by other bounds, changing $m_N$ or the Yukawa couplings has very little effects. One can see that $m_L$ is excluded up to $\sim 149$, 170 and 187 GeV for $SU(2)$, $SU(3)$ and $SU(4)$ respectively. These numbers are lowered by the fact that Ref. \cite{Aaboud:2016yus} contains a considerable upper fluctuation in its number of events. As a reference point, we mention what the limits would have been with the full 36 fb$^{-1}$ of integrated luminosity currently available at 13 TeV. For this, we naively scale all signals, numbers of events, backgrounds and their uncertainties by the ratio of the full to current integrated luminosity. This procedure can exclude masses as high as $\sim 166$, 191 and 207 GeV for $SU(2)$, $SU(3)$ and $SU(4)$ respectively.\footnote{Technically, there would be a narrow band at lower mass that would not be excluded as the statistical fluctuation mentioned above would become very important with 36 fb$^{-1}$ of integrated luminosity.}

The bounds we obtain correspond to what should have been expected. Indeed, these events are similar to electroweakino production, with similar cross sections but considerably less MET. One would therefore expect the resulting limits to be weaker, but still considerably above the mass of the $Z$ and $W$ bosons. Several electroweakino searches look for production of a neutralino $\chi_2$ and a chargino $\chi^+_1$, with the chargino and heavy neutralino decaying to massive gauge bosons and the lightest neutralino $\chi_1$. Ref. \cite{CMS:2017fdz} finds that, for light $\chi_1$ and degenerate $\chi_2$ and $\chi^+_1$, the heavy neutralino and the chargino are forced to be above 450 GeV. Our results are therefore reasonable.

\subsubsection{$\tilde{\eta}$ decaying inside the detector}\label{ssSec:ColliderInside}
The meson $\tilde{\eta}$ decaying inside the detector represents the most experimentally challenging case. Indeed, what results from each $n_1$ is an emerging jet, which will neither be properly reconstructed as a jet or taken into account in the MET. The only usable signature is then the leptons coming from the decay of massive gauge bosons.

The strategy that we adopt for $\tilde{\eta}$ decaying inside the detector is simply the same as if it were to decay promptly, but without requiring the presence of two jets. Limits will therefore be considerably weaker. The simulation and analysis procedure is the same as in Sec.~\ref{ssSec:ColliderPrompt}, without of course the cut on the number of jets. We also include the searches of Refs. \cite{Aad:2016ett} and \cite{Khachatryan:2016poo}, as the missing numbers are irrelevant when jets are not counted. Fig.~\ref{fig:CLsEmerging} shows the $CL_s$ as a function of $m_L$ for small Yukawa couplings and $m_N$. The mass $m_L$ is excluded up to $\sim 134$, 148 and 160 GeV for $SU(2)$, $SU(3)$ and $SU(4)$ respectively.

\subsubsection{$\tilde{\eta}$ decaying outside the detector}\label{ssSec:ColliderOutside}
In contrast to $\tilde{\eta}$ decaying inside the detector, $\tilde{\eta}$ decaying outside the detector is very easy to constrain. The transverse energy associated to $n_1$ is simply transmitted to MET. Supersymmetry searches for electroweakinos can then be applied directly.

\begin{figure}
  \begin{center}
   \includegraphics[width=0.57\textwidth, viewport = 0 0 450 366]{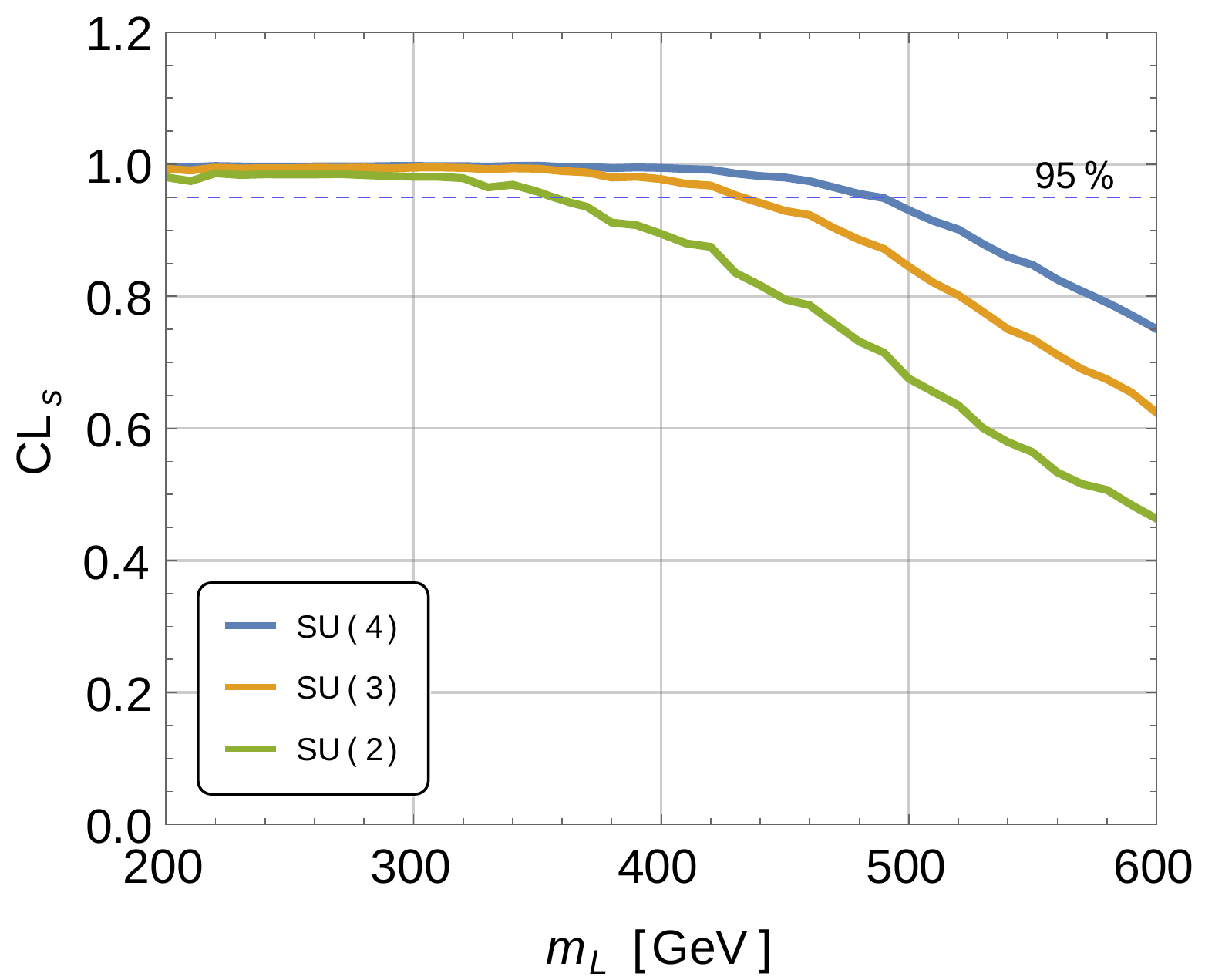}
   \caption{$CL_s$ for stable $n_1$ as a function of $m_L$ and for different $SU(N)$. The mass $m_N$ is set to 5 GeV and the Yukawa couplings to $y=\tilde{y}=0.1$.}
   \label{fig:CLsStable}
  \end{center}
\end{figure}

The search we focus on is the previously mentioned Ref. \cite{CMS:2017fdz}. It is the strongest search presently available that constrains the production of $\chi_2$ and $\chi^+_1$, with these gauginos decaying to massive gauge bosons and $\chi_1$. We implemented a code to simulate this search and verified that we could reproduce their results with good accuracy. Events are simulated as in Sec.~\ref{ssSec:ColliderPrompt}, but with $n_1$ now added to the MET. Fig.~\ref{fig:CLsStable} shows the $CL_s$ as a function of $m_L$ for small Yukawa couplings and $m_N$. The mass $m_L$ is excluded up to $\sim 355$, 430, 490 GeV for $SU(2)$, $SU(3)$ and $SU(4)$ respectively.

\section{Vacuum stability}\label{Sec:Vacuum}
We now discuss the effect of the additional fermions on vacuum stability. These new fermions inherently need to communicate with the Higgs for the relaxion mechanism to work. This in turn affects the running of the Higgs quartic. If at a given scale the Higgs quartic reaches a value that is too negative, metastability will become an issue. This is not a constraint in itself, as new physics that modifies the running of the quartic can always be postulated to come into play before that scale. It however gives an upper bound on the validity of the theory that can potentially be smaller than the cutoff of Sec.~\ref{sSec:RelaxionExplanation}. To evaluate this bound, we proceed to study how the running of the Higgs quartic is modified by the new sector.

We first list the beta functions of the most important couplings. The couplings that are evolved are: all gauge couplings, $y$, $\tilde{y}$ and the top Yukawa coupling $y_t$. As we are ultimately interested in an estimate of the scale where new physics must come into play, we limit ourselves to leading order in the beta function. All beta functions were verified with SARAH \cite{Staub:2013tta}. Assuming the group $\mathcal{G}$ is $SU(N)$, the beta function of the gauge couplings are
\begin{equation}\label{eq:BetaGauge}
  \begin{aligned}
    \beta(g')   &=  \frac{41+4N}{6} \frac{g'^3}{16\pi^2},\qquad 
    &\beta(g)   &=  \frac{-19+4N}{6}\frac{g^3}{16\pi^2},\\
    \beta(g_s)  &= -7 \frac{g_s^3}{16\pi^2}, 
    &\beta(g_f) &=  \frac{6-11N}{3} \frac{g_f^3}{16\pi^2},
  \end{aligned}
\end{equation}
where $g_f$ is the gauge coupling constant of $\mathcal{G}$. The beta function of the top Yukawa is
\begin{equation}\label{eq:Betayt}
  \beta(y_t)=\frac{y_t}{16^2}\left[-8g_s^2 + \frac{9}{2}y_t^2 - \frac{9}{4} g^2 - \frac{17}{12}g'^2 + N y^2 + N \tilde{y}^2\right],
\end{equation}
while those of $y$ and $\tilde{y}$ are
\begin{equation}\label{eq:Betayytilde}
  \begin{aligned}
    \beta(y) &=\frac{y}{16\pi^2}\left[-\frac{3}{4}g'^2-\frac{9}{4}g^2-\frac{3(N^2-1)}{N}g_f^2+\frac{2N+3}{2}y^2+N\tilde{y}^2 + 3y_t^2\right],\\
    \beta(\tilde{y}) &=\frac{\tilde{y}}{16\pi^2}\left[-\frac{3}{4}g'^2-\frac{9}{4}g^2-\frac{3(N^2-1)}{N}g_f^2+\frac{2N+3}{2}\tilde{y}^2+Ny^2+3y_t^2\right].
  \end{aligned}
\end{equation}
Finally, the beta function of the Higgs quartic is given by
\begin{equation}\label{eq:BetaQuartic}
  \begin{aligned}
    \beta(\lambda) &= \frac{1}{16\pi^2}\left[24\lambda^2-3g'^2\lambda-9g^2\lambda+\frac{3}{8}g'^4+\frac{3}{4}g^2g'^2+\frac{9}{8}g^4-6y_t^4\right.\\
                   &\hspace{1.5cm}\left.+12y_t^2\lambda -2Ny^4-2N\tilde{y}^4+4Ny^2\lambda+4N\tilde{y}^2\lambda\right].
  \end{aligned}
\end{equation}
We note that the beta function of the quartic depends on the fourth power of $y$ and $\tilde{y}$. As such, the Higgs quartic will be almost unaffected for small $y$ and $\tilde{y}$, but will have its behavior drastically altered as these coupling approach 1.

\begin{figure}
  \begin{center}
   \includegraphics[width=0.57\textwidth, viewport = 0 0 450 360]{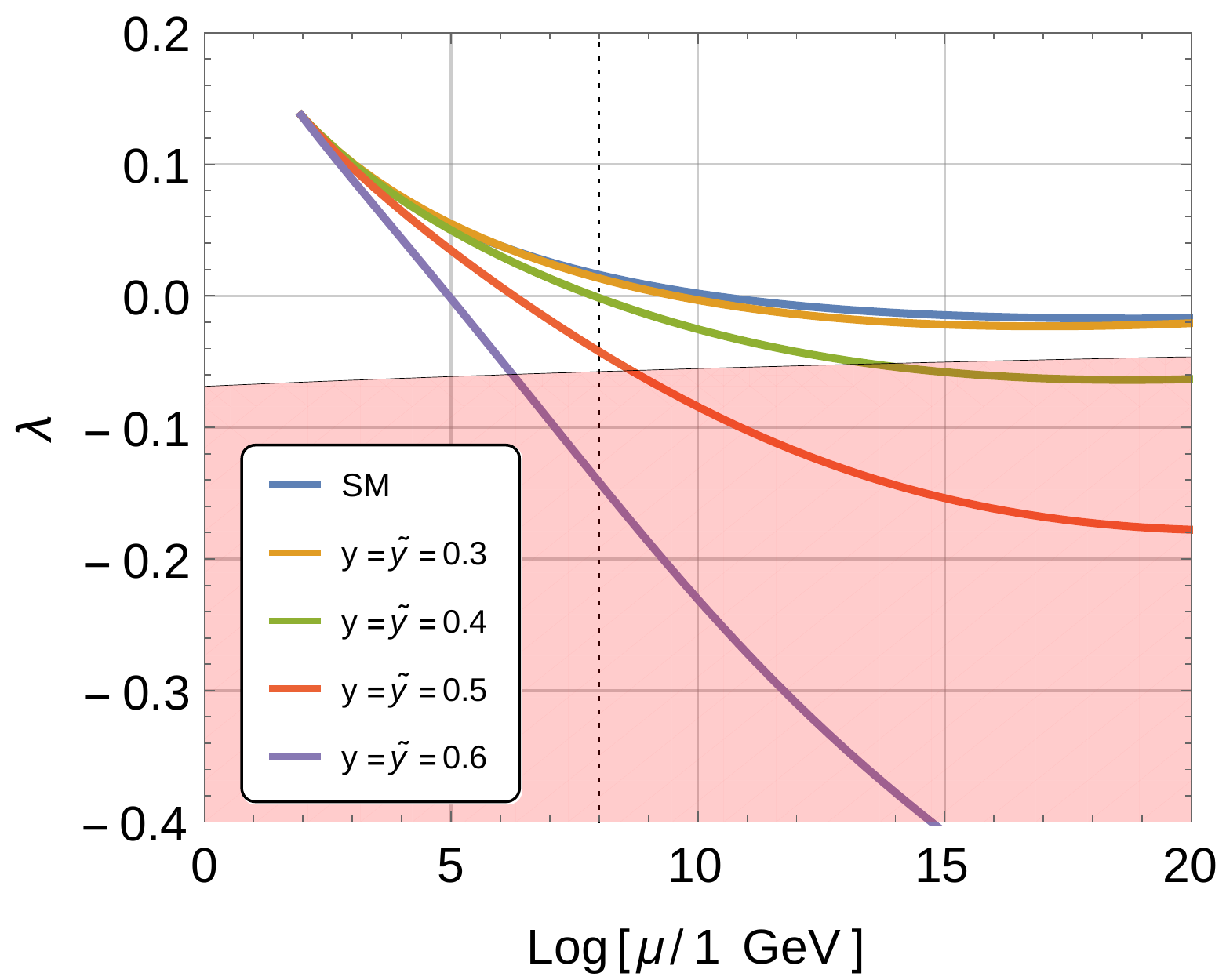}
   \caption{Evolution of the Higgs quartic for different values of $y=\tilde{y}$. The coupling $g_f$ is set to 1 at $m_Z$ and $\mathcal{G}$ to $SU(3)$. The pink region corresponds to the instability bound. The dotted line corresponds to the maximum cutoff of $10^8$~GeV.}
   \label{fig:VacuumStability1}
  \end{center}
\end{figure}

The parameters are evolved starting from the $Z$ boson mass $m_Z$  using the $\overline{MS}$ values from Ref.~\cite{Degrassi:2012ry}. Fig.~\ref{fig:VacuumStability1} shows an example of the evolution of the quartic for $\mathcal{G} = SU(3)$. The instability bound is taken from Ref.~\cite{Isidori:2001bm} (see also Ref.~\cite{Bandyopadhyay:2016oif}) and corresponds to a lifetime of the metastable vacuum shorter than the age of the Universe. As can be seen, Yukawa couplings of more than $\sim 0.4$ require additional new physics below the Planck scale. However, the only regions where metastability requires new physics below the estimated cutoff of $10^8$~GeV correspond to values of the Yukawa couplings of $\gtrsim 0.5$, which are already ruled out by bounds from EWPT and Higgs decay for reasonable values of $m_L$ and $m_N$.

\section{Tuning}\label{Sec:Tuning}
We now proceed to discuss a certain difficulty that arises as the confinement scale decreases which makes the relaxion mechanism less likely to function properly. Its success will require the different parameters to satisfy a specific quasi-equality. We associate to this a measure of what we refer to as tuning. Any additional tunings related to the axion or other cosmological considerations are ignored. We instead refer to previous work for this (see for example Refs. \cite{Gupta:2015uea, DiChiara:2015euo, Choi:2016luu, Flacke:2016szy}).

\begin{figure}
  \begin{center}
   \includegraphics[width=0.57\textwidth, viewport = 0 0 360 350]{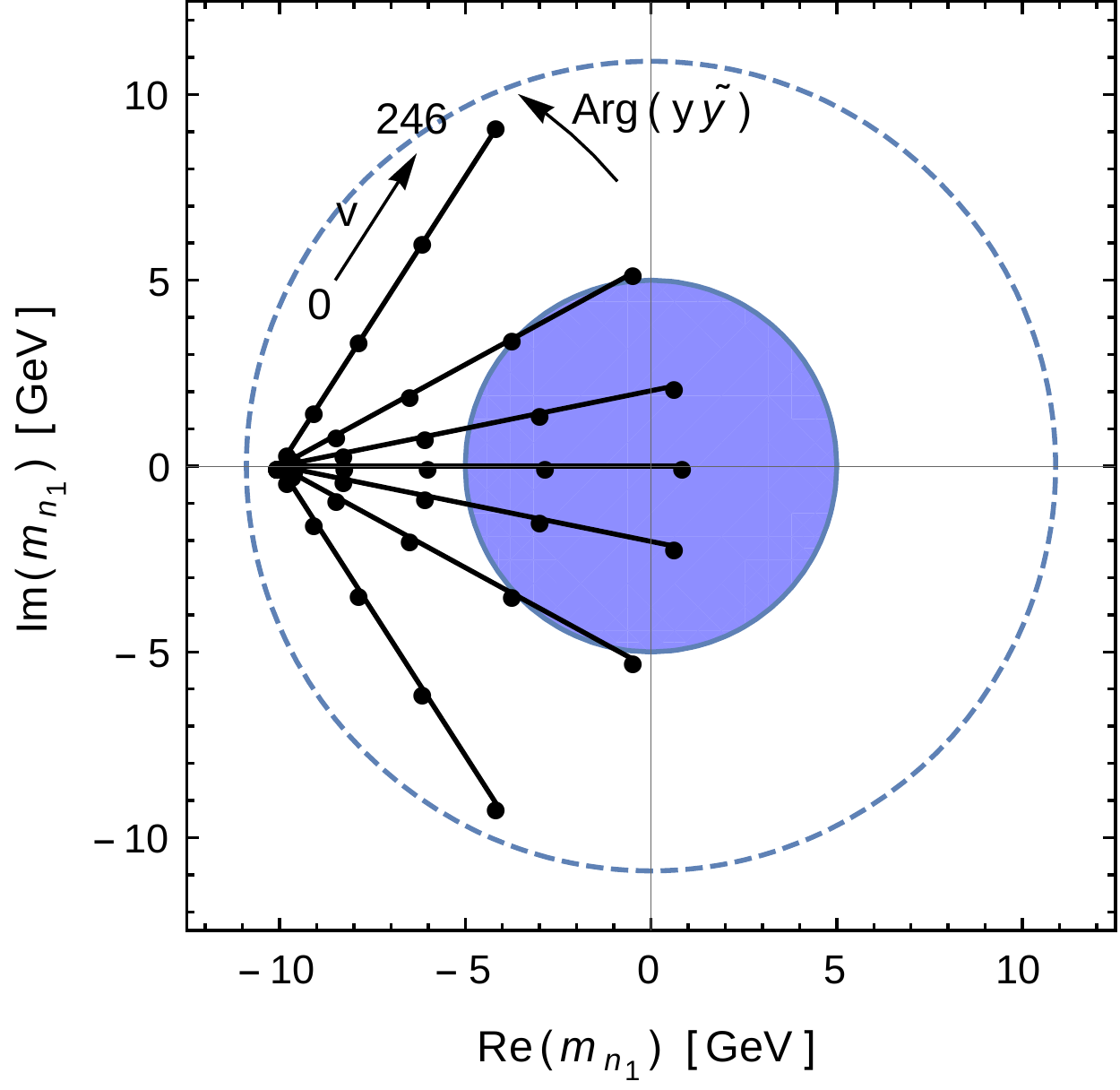}
   \caption{Evolution of $m_{n_1}$ as $v$ increases. Different curves correspond to different arguments of $y\tilde{y}$. Other parameters are set to: $\Lambda=10$ GeV, $m_L=250$ GeV, $m_N=-10$ GeV, $|\tilde{y}|=0.3$ and $|y|=0.3$. The blue circle represents the region where $|m_{n_1}|$ is below the confinement scale. The region inside the dashed line satisfies $c_0<c_2v^2$. The black points represent respectively $v=\{0,50,100,150,200,246\}$ GeV.}
   \label{fig:mN1}
  \end{center}
\end{figure}

As was mentioned in Sec.~\ref{sSec:RelaxionExplanation}, the mass of $n_1$ is given by
\begin{equation}\label{eq:mn1}
  m_{n_1}\approx m_N+\frac{y\tilde{y}}{2}\frac{v^2}{m_L},
\end{equation}
which is true even for complex parameters. For low $\Lambda$ and large enough Yukawa couplings, the second term of Eq.~(\ref{eq:mn1}) can be considerably higher than the confinement scale. The relaxion mechanism can still technically work, as long as there is a partial cancellation with $m_N$ such that $m_{n_1}$ is smaller than $\Lambda$. This can at first seem like a tuning, but the situation is slightly more complicated. Indeed, the quasi-equality in magnitude of the two terms in Eq.~(\ref{eq:mn1}) can be explained dynamically. When the Higgs first acquires an expectation value, $m_{n_1}$ is considerably higher than the confinement scale. There is therefore no back-reaction potential and $m_{n_1}$ is scanned without anything stopping it. Assuming all Yukawa couplings and $m_L$ to be real and positive and $m_N$ to be negative, $m_{n_1}$ would eventually become smaller than the confinement scale and the back-reaction term would form. This could then explain dynamically the necessary quasi-equality of the two terms of Eq.~(\ref{eq:mn1}). However, this is only true as long as all parameters are real, for which there is no guarantee. In the general case, the scanning of $m_{n_1}$ can go in any direction in the complex plane starting from $m_N$. The curve representing the evolution of $m_{n_1}$ would have to intercept a small circle of radius $\Lambda$ around the origin for the relaxion mechanism to work. This is where the tuning comes from. In the hope of making the argument clearer, Fig.~\ref{fig:mN1} shows the evolution of $m_{n_1}$ for different phases of $y\tilde{y}$. As can be seen, $|m_{n_1}|$ only drops below the condensation scale for small phases. Assuming any scanning direction to be as likely, the tuning can be defined as the fraction of directions that lead to the relaxion mechanism working, which is
\begin{equation}\label{eq:Delta}
  \text{Tuning} = \frac{2\Lambda}{2\pi |m_N|}\approx\left|\frac{2m_L\Lambda}{\pi y\tilde{y} v^2}\right|.
\end{equation}
This is a tuning in the same sense as the strong $CP$-problem.

We mention that assuming all parameters real would restore $CP$-symmetry in the new strongly coupled sector. Of course, $CP$-violation would be reintroduced in that sector via interactions with the Standard Model. The phases could then be potentially suppressed, which would make the dynamical selection mechanism we explained before more likely to succeed. Our tuning is then a tuning in the sense that there are no reasons in general for the parameters to be real.

As a side note, we mention that the regions of parameter space that are very fine-tuned have at least the advantage of providing a solution to the coincidence problem of requiring $m_L$ and $m_N$ close to $v$. Indeed, these regions are characterized by the back-reaction potential only appearing when $|m_{n_1}|$ drops below $\Lambda$ as $m_{n_1}$ evolves along its trajectory. This leads to $|m_{N}|\approx |y\tilde{y}v^2/2m_L|$, or alternatively $v\approx |2m_N m_L/y\tilde{y}|^{1/2}$. The vacuum expectation value being close to $m_N$ and $m_L$ is then explained dynamically, though of course at the price of some tuning.

\section{Summary of the constraints}\label{Sec:SummaryBounds}
We now assemble together all the bounds derived in the Sec.~\ref{Sec:Constraints} and present representative plots including the tuning of Sec.~\ref{Sec:Tuning}. To reduce the dimension of the parameter space to a manageable number, we make a few simplifying assumptions. First, we assume $y=2\tilde{y}$. Second, we take the group $\mathcal{G}$ to be $SU(3)$. Third, for the relaxion mechanism to function properly, $m_N$ must be such that $m_{n_1}$ is smaller than the confinement scale $\Lambda$ and that $|c_0|$ is inferior to $|c_2v^2|$ (see Eq.~(\ref{eq:rho})). With this in mind, we set $m_N=-y \tilde{y}v^2/2m_L$.\footnote{We neglect the effect of subleading corrections as these would have very little impact on the experimental constraints. We also add a correction of 1~GeV to $m_{n_1}$ during collider simulation to avoid issues with decay in \texttt{PYTHIA}. This correction has negligible effect on the efficiencies.} This has the advantage of satisfying both conditions, while providing the maximal value of $|m_N|$ allowed. This parametrization also minimizes any fine-tuning that might be introduced by having $m_N$ too small. In a certain way, this choice of $m_N$ can be seen as the best case scenario. In comparison, keeping $m_N$ fixed in a scan would lead to both regions that are already ruled out by virtue of the relaxion mechanism not working and regions where $m_N$ is very fine-tuned. We also note that the exact value of $m_N$ has very little impact on any of the experimental constraints as long as it is sufficiently small, which has to be the case for Yukawa values that are not already ruled out by EWPT and Higgs branching ratio measurements. Under these assumptions, the three free parameters left are therefore $m_L$, $y=2\tilde{y}$ and the confinement scale $\Lambda$. We scan over the first two and fix $\Lambda$ to different representative values.

\begin{figure}
  \centering
  \begin{subfigure}[b]{0.5\textwidth}
    \centering
    \caption{$\Lambda=$ 10 MeV}
    \includegraphics[width=0.93\linewidth, viewport = 0 -20 450 430]{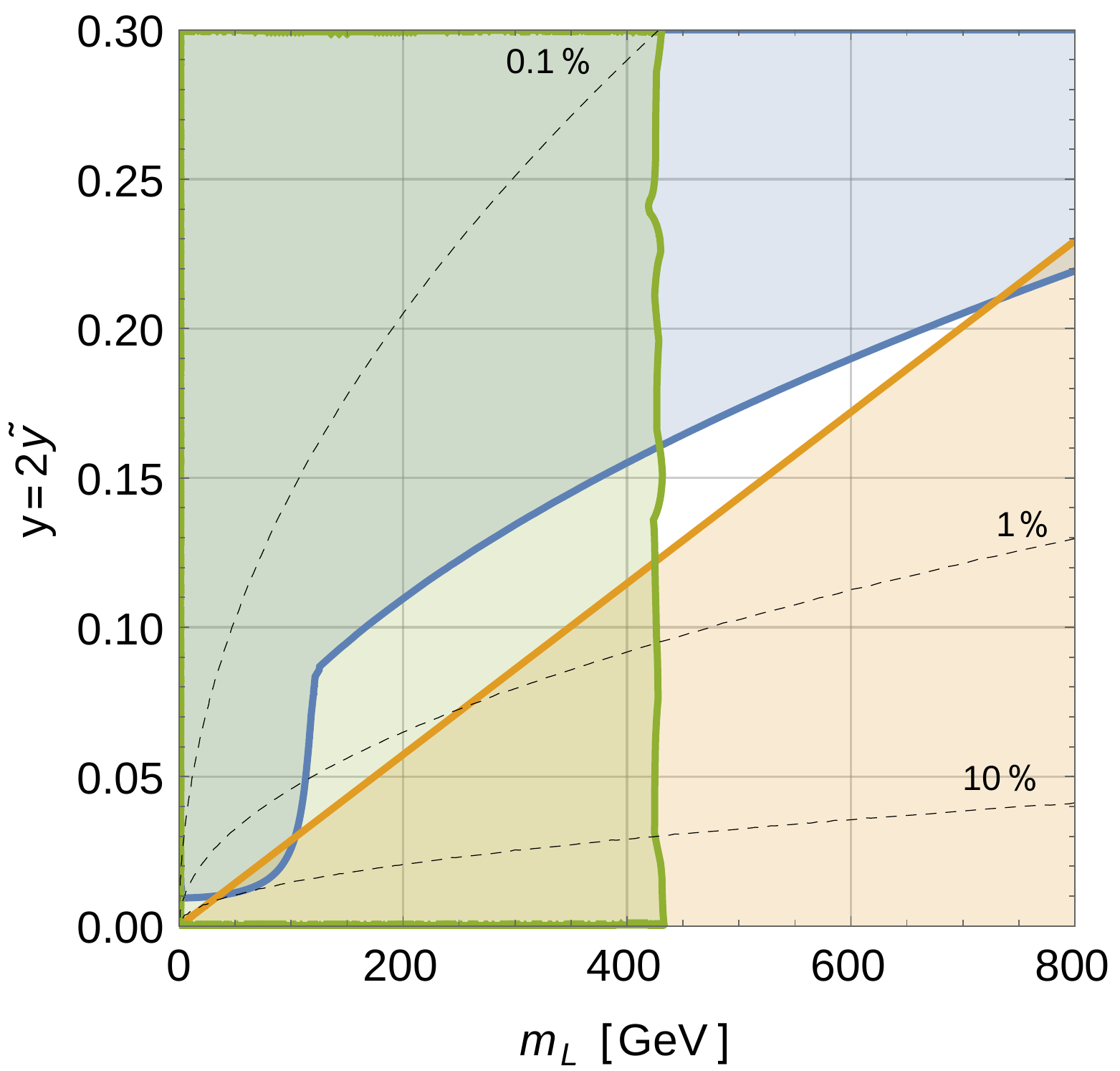}
    \label{fig:S1}
  \end{subfigure}%
    ~\quad
  \begin{subfigure}[b]{0.5\textwidth}
    \centering
    \caption{$\Lambda=$ 300 MeV}
    \includegraphics[width=0.93\linewidth, viewport = 0 -20 450 430]{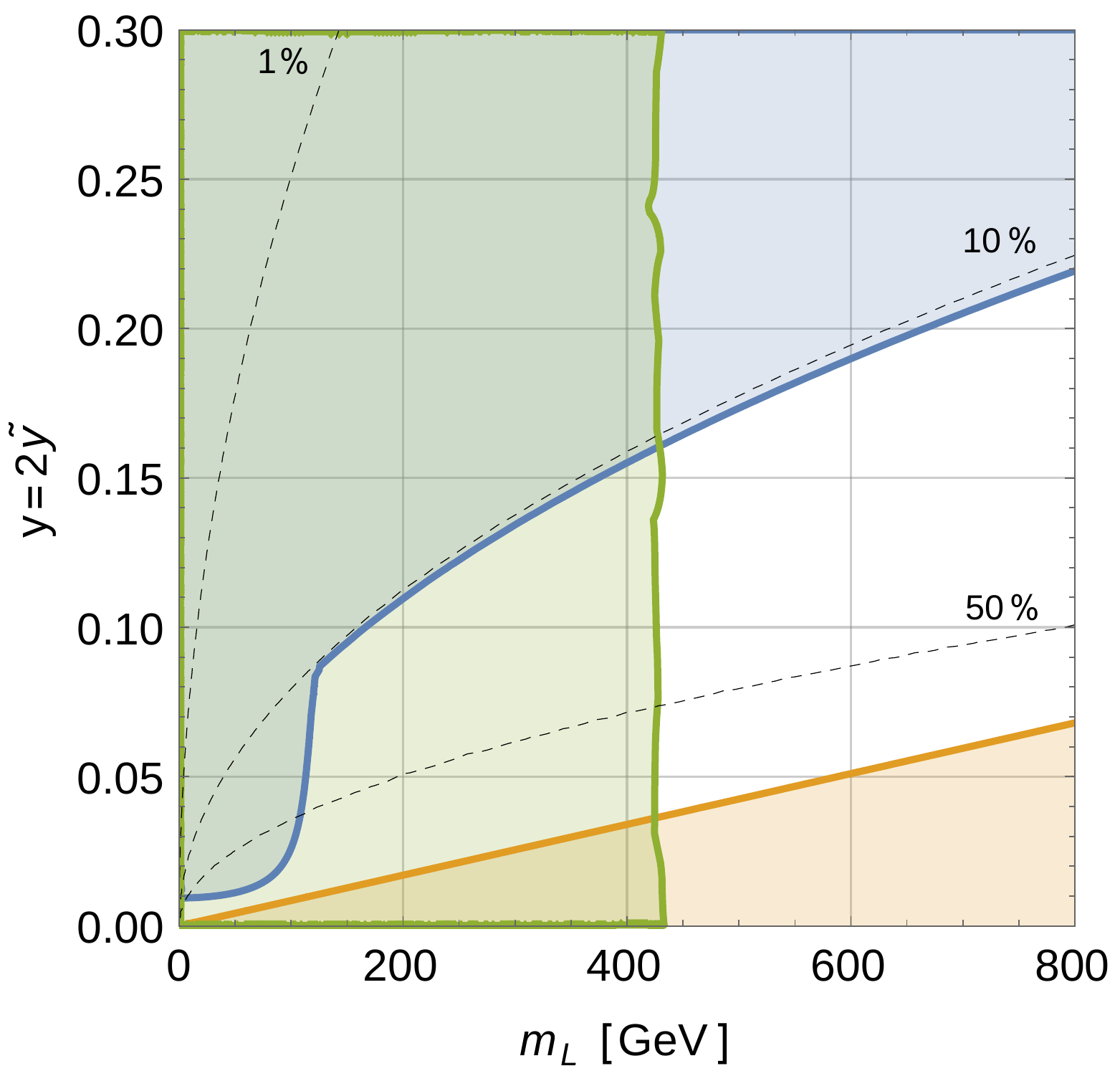}
    \label{fig:S2}
  \end{subfigure}
  \begin{subfigure}[b]{0.5\textwidth}
    \centering
    \caption{$\Lambda=$ 10 GeV}
    \includegraphics[width=0.93\linewidth, viewport = 0 -20 450 430]{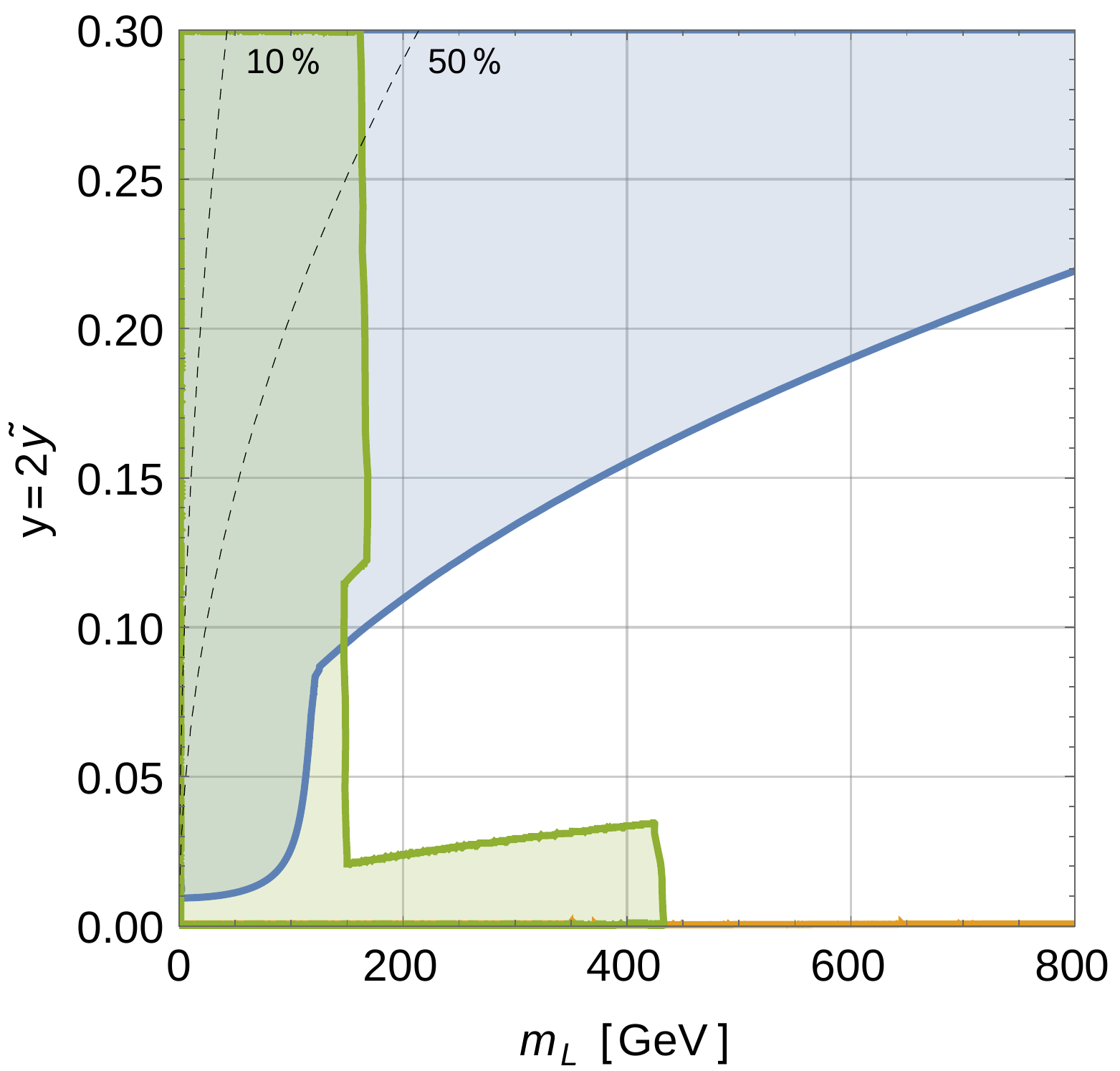}
    \label{fig:S3}
  \end{subfigure}%
    ~\quad
  \begin{subfigure}[b]{0.5\textwidth}
    \centering
    \caption{$\Lambda=$ 25 GeV}
    \includegraphics[width=0.93\linewidth, viewport = 0 -20 450 430]{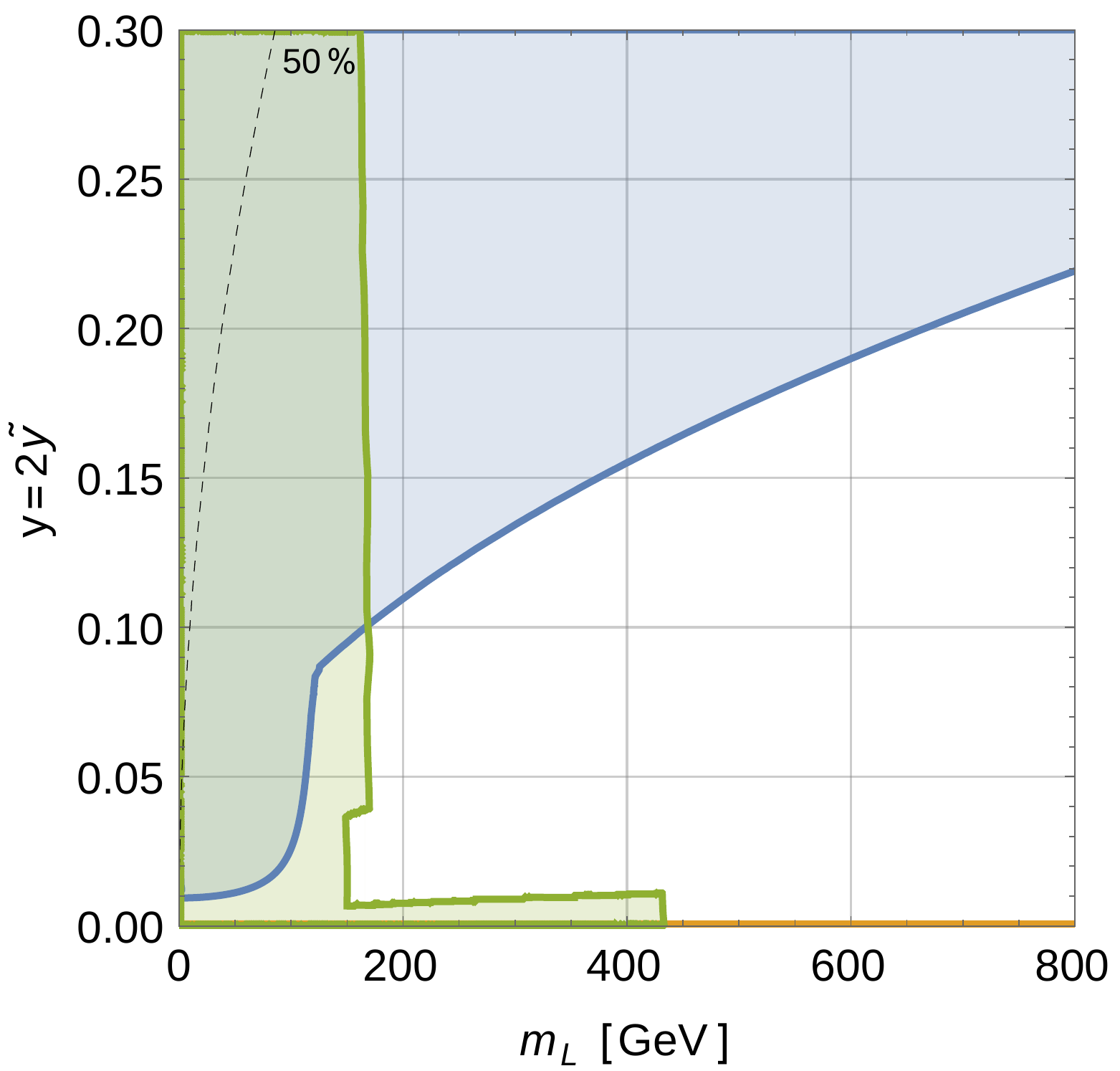}
    \label{fig:S4}
  \end{subfigure}
  \caption{Summary of the experimental constraints at 95$\%$ CL for different values of $\Lambda$. The green, blue and orange regions are ruled out respectively by direct collider, Higgs branching ratio and BBN constraints. The dashed curves represent tuning.}\label{fig:SummaryPlot}
\end{figure}

The results are shown in Fig.~\ref{fig:SummaryPlot}. 
A few comments are in order. The four benchmarks for the confinement scale are chosen to represent specific behaviors. The values of 10~MeV and 25~GeV represent respectively the minimal value for which the relaxion mechanism is still relevant and the maximum value for which our approximations are still reasonably valid. The benchmark of 300~MeV is chosen to be slightly above the minimum value such that the decay to two charged pions is possible. The value of 10~GeV represents a region where all three types of collider constraints are relevant, with the bounds from stable $\tilde{\eta}$'s, emerging jets and promptly decaying $\tilde{\eta}$'s being respectively applicable to small, intermediary and larger Yukawa couplings. The different colored regions are ruled out by the constraints of Sec.~\ref{Sec:Constraints}. All bounds correspond to 95$\%$ CL. We do not include the bounds from EWPT as these regions are already excluded by the Higgs branching ratio and collider constraints. The bounds on chargino from LEP would also not contribute any additional constraints \cite{LEP:ch1, LEP:ch2}. Collider bounds are applied using the three cases of Sec.~\ref{sSec:Collider}. If the decay length of $\tilde{\eta}$ is less than 1 mm, it is considered to decay promptly. If its decay length is between 1 mm and 1 m, it is considered to decay inside the detector. For longer decay lengths, it is considered to decay outside the detector. 

The dashed contours represent the tuning discussed in Sec.~\ref{Sec:Tuning}. As can be seen in Fig.~\ref{fig:SummaryPlot}, the tuning can become quite severe for low enough confinement scales. BBN is crucial, as it prevents the mechanism from taking refuge from tuning in the low Yukawa couplings region. For a confinement scale of 10~MeV, the tuning can be at best at the percent level.

\section{Conclusions}\label{Sec:Conclusion}
As the LHC continues to probe higher and higher energies, BSM theories that seek to stabilize the electroweak scale with colored partners are becoming increasingly under tension. In contrast to these models, the relaxion mechanism attempts to explain the smallness of the electroweak scale via a dynamical selection process, thereby bypassing the need for any colored partners. The relaxion field scans over the mass square of the Higgs and a back-reaction potential stops its evolution shortly after the Higgs acquires an expectation value. This back-reaction potential can be generated in several ways. One of them is via the introduction of new strongly coupled vector-fermions.

The goal of this article was to study the constraints on such vector-fermions and investigate any potential tuning. More specifically, we looked at the non-QCD model of Ref.~\cite{Graham:2015cka} for small confinement scales. We investigated bounds coming from EWPT, Higgs decay, BBN and direct collider searches. Additionally, we studied the impact of such fermions on vacuum stability and found that some regions of parameter space require new physics below the previously estimated cutoff. Combining all of these constraints, we showed that the parameter space becomes increasingly restricted and tuned as the confinement scale decreases. For example, a confinement scale of a few tens of MeVs would require tuning at the percent level.

The conclusion that the confinement scale cannot be too small without introducing fine-tuning is probably the most important lesson from our article. The confinement scale also cannot be too high without reintroducing fine-tuning in the back-reaction potential. Additionally, this region has been studied in Refs.~\cite{Antipin:2015jia, Agugliaro:2016clv}. This leaves the region close to the electroweak scale as the most natural choice for the confinement scale. This region incidentally has not been the subject of any dedicated study. If the relaxion is ever to become a solution to the hierarchy problem on par with Supersymmetry and composite Higgs, a dedicated study of this scenario should seriously be considered.

\acknowledgments
This work was supported by Funda\c{c}\~{a}o de Amparo \`{a} Pesquisa do Estado de S\~{a}o Paulo (FAPESP) and Conselho Nacional de Ci\^{e}ncia e Tecnologia (CNPq). We would also like to thank Chee Sheng Fong for useful discussions.

\bibliography{biblio}

\providecommand{\href}[2]{#2}\begingroup\raggedright\begin{thebibliography}{100}

\bibitem{Golfand:1971iw}
{\relax Yu}.~A. Golfand and E.~P. Likhtman, {\it {Extension of the Algebra of
  Poincare Group Generators and Violation of p Invariance}},  {\em JETP Lett.}
  {\bf 13} (1971) 323--326. [Pisma Zh. Eksp. Teor. Fiz.13,452(1971)].

\bibitem{Volkov:1973ix}
D.~V. Volkov and V.~P. Akulov, {\it {Is the Neutrino a Goldstone Particle?}},
  {\em Phys. Lett.} {\bf 46B} (1973) 109--110.

\bibitem{Wess:1974tw}
J.~Wess and B.~Zumino, {\it {Supergauge Transformations in Four-Dimensions}},
  {\em Nucl. Phys.} {\bf B70} (1974) 39--50.

\bibitem{Wess:1974jb}
J.~Wess and B.~Zumino, {\it {Supergauge Invariant Extension of Quantum
  Electrodynamics}},  {\em Nucl. Phys.} {\bf B78} (1974) 1.

\bibitem{Ferrara:1974pu}
S.~Ferrara and B.~Zumino, {\it {Supergauge Invariant Yang-Mills Theories}},
  {\em Nucl. Phys.} {\bf B79} (1974) 413.

\bibitem{Dimopoulos:1981zb}
S.~Dimopoulos and H.~Georgi, {\it {Softly Broken Supersymmetry and SU(5)}},
  {\em Nucl. Phys.} {\bf B193} (1981) 150--162.

\bibitem{Bellazzini:2014yua}
B.~Bellazzini, C.~Cs{\'a}ki, and J.~Serra, {\it {Composite Higgses}},  {\em
  Eur. Phys. J.} {\bf C74} (2014), no.~5 2766,
  [\href{http://arxiv.org/abs/1401.2457}{{\tt arXiv:1401.2457}}].

\bibitem{Panico:2015jxa}
G.~Panico and A.~Wulzer, {\it {The Composite Nambu-Goldstone Higgs}},  {\em
  Lect. Notes Phys.} {\bf 913} (2016) pp.1--316,
  [\href{http://arxiv.org/abs/1506.01961}{{\tt arXiv:1506.01961}}].

\bibitem{Chacko:2005pe}
Z.~Chacko, H.-S. Goh, and R.~Harnik, {\it {The Twin Higgs: Natural electroweak
  breaking from mirror symmetry}},  {\em Phys. Rev. Lett.} {\bf 96} (2006)
  231802, [\href{http://arxiv.org/abs/hep-ph/0506256}{{\tt hep-ph/0506256}}].

\bibitem{Burdman:2006tz}
G.~Burdman, Z.~Chacko, H.-S. Goh, and R.~Harnik, {\it {Folded supersymmetry and
  the LEP paradox}},  {\em JHEP} {\bf 02} (2007) 009,
  [\href{http://arxiv.org/abs/hep-ph/0609152}{{\tt hep-ph/0609152}}].

\bibitem{Graham:2015cka}
P.~W. Graham, D.~E. Kaplan, and S.~Rajendran, {\it {Cosmological Relaxation of
  the Electroweak Scale}},  {\em Phys. Rev. Lett.} {\bf 115} (2015), no.~22
  221801, [\href{http://arxiv.org/abs/1504.07551}{{\tt arXiv:1504.07551}}].

\bibitem{Espinosa:2015eda}
J.~R. Espinosa, C.~Grojean, G.~Panico, A.~Pomarol, O.~Pujol{\`a}s, and
  G.~Servant, {\it {Cosmological Higgs-Axion Interplay for a Naturally Small
  Electroweak Scale}},  {\em Phys. Rev. Lett.} {\bf 115} (2015), no.~25 251803,
  [\href{http://arxiv.org/abs/1506.09217}{{\tt arXiv:1506.09217}}].

\bibitem{Hardy:2015laa}
E.~Hardy, {\it {Electroweak relaxation from finite temperature}},  {\em JHEP}
  {\bf 11} (2015) 077, [\href{http://arxiv.org/abs/1507.07525}{{\tt
  arXiv:1507.07525}}].

\bibitem{Patil:2015oxa}
S.~P. Patil and P.~Schwaller, {\it {Relaxing the Electroweak Scale: the Role of
  Broken dS Symmetry}},  {\em JHEP} {\bf 02} (2016) 077,
  [\href{http://arxiv.org/abs/1507.08649}{{\tt arXiv:1507.08649}}].

\bibitem{Antipin:2015jia}
O.~Antipin and M.~Redi, {\it {The Half-composite Two Higgs Doublet Model and
  the Relaxion}},  {\em JHEP} {\bf 12} (2015) 031,
  [\href{http://arxiv.org/abs/1508.01112}{{\tt arXiv:1508.01112}}].

\bibitem{Jaeckel:2015txa}
J.~Jaeckel, V.~M. Mehta, and L.~T. Witkowski, {\it {Musings on cosmological
  relaxation and the hierarchy problem}},  {\em Phys. Rev.} {\bf D93} (2016),
  no.~6 063522, [\href{http://arxiv.org/abs/1508.03321}{{\tt
  arXiv:1508.03321}}].

\bibitem{Gupta:2015uea}
R.~S. Gupta, Z.~Komargodski, G.~Perez, and L.~Ubaldi, {\it {Is the Relaxion an
  Axion?}},  {\em JHEP} {\bf 02} (2016) 166,
  [\href{http://arxiv.org/abs/1509.00047}{{\tt arXiv:1509.00047}}].

\bibitem{Batell:2015fma}
B.~Batell, G.~F. Giudice, and M.~McCullough, {\it {Natural Heavy
  Supersymmetry}},  {\em JHEP} {\bf 12} (2015) 162,
  [\href{http://arxiv.org/abs/1509.00834}{{\tt arXiv:1509.00834}}].

\bibitem{Matsedonskyi:2015xta}
O.~Matsedonskyi, {\it {Mirror Cosmological Relaxation of the Electroweak
  Scale}},  {\em JHEP} {\bf 01} (2016) 063,
  [\href{http://arxiv.org/abs/1509.03583}{{\tt arXiv:1509.03583}}].

\bibitem{Marzola:2015dia}
L.~Marzola and M.~Raidal, {\it {Natural relaxation}},  {\em Mod. Phys. Lett.}
  {\bf A31} (2016) 1650215, [\href{http://arxiv.org/abs/1510.00710}{{\tt
  arXiv:1510.00710}}].

\bibitem{Choi:2015fiu}
K.~Choi and S.~H. Im, {\it {Realizing the relaxion from multiple axions and its
  UV completion with high scale supersymmetry}},  {\em JHEP} {\bf 01} (2016)
  149, [\href{http://arxiv.org/abs/1511.00132}{{\tt arXiv:1511.00132}}].

\bibitem{Kaplan:2015fuy}
D.~E. Kaplan and R.~Rattazzi, {\it {Large field excursions and approximate
  discrete symmetries from a clockwork axion}},  {\em Phys. Rev.} {\bf D93}
  (2016), no.~8 085007, [\href{http://arxiv.org/abs/1511.01827}{{\tt
  arXiv:1511.01827}}].

\bibitem{DiChiara:2015euo}
S.~Di~Chiara, K.~Kannike, L.~Marzola, A.~Racioppi, M.~Raidal, and C.~Spethmann,
  {\it {Relaxion Cosmology and the Price of Fine-Tuning}},  {\em Phys. Rev.}
  {\bf D93} (2016), no.~10 103527, [\href{http://arxiv.org/abs/1511.02858}{{\tt
  arXiv:1511.02858}}].

\bibitem{Ibanez:2015fcv}
L.~E. Ibanez, M.~Montero, A.~Uranga, and I.~Valenzuela, {\it {Relaxion
  Monodromy and the Weak Gravity Conjecture}},  {\em JHEP} {\bf 04} (2016) 020,
  [\href{http://arxiv.org/abs/1512.00025}{{\tt arXiv:1512.00025}}].

\bibitem{Fonseca:2016eoo}
N.~Fonseca, L.~de~Lima, C.~S. Machado, and R.~D. Matheus, {\it {Large field
  excursions from a few site relaxion model}},  {\em Phys. Rev.} {\bf D94}
  (2016), no.~1 015010, [\href{http://arxiv.org/abs/1601.07183}{{\tt
  arXiv:1601.07183}}].

\bibitem{Gertov:2016uzs}
H.~Gertov, F.~Sannino, L.~Pearce, and L.~Yang, {\it {Baryogenesis via
  Elementary Goldstone Higgs Relaxation}},  {\em Phys. Rev.} {\bf D93} (2016),
  no.~11 115042, [\href{http://arxiv.org/abs/1601.07753}{{\tt
  arXiv:1601.07753}}].

\bibitem{Fowlie:2016jlx}
A.~Fowlie, C.~Balazs, G.~White, L.~Marzola, and M.~Raidal, {\it {Naturalness of
  the relaxion mechanism}},  {\em JHEP} {\bf 08} (2016) 100,
  [\href{http://arxiv.org/abs/1602.03889}{{\tt arXiv:1602.03889}}].

\bibitem{Evans:2016htp}
J.~L. Evans, T.~Gherghetta, N.~Nagata, and Z.~Thomas, {\it {Naturalizing
  Supersymmetry with a Two-Field Relaxion Mechanism}},  {\em JHEP} {\bf 09}
  (2016) 150, [\href{http://arxiv.org/abs/1602.04812}{{\tt arXiv:1602.04812}}].

\bibitem{Kobayashi:2016bue}
T.~Kobayashi, O.~Seto, T.~Shimomura, and Y.~Urakawa, {\it {Relaxion window}},
  \href{http://arxiv.org/abs/1605.06908}{{\tt arXiv:1605.06908}}.

\bibitem{Hook:2016mqo}
A.~Hook and G.~Marques-Tavares, {\it {Relaxation from particle production}},
  {\em JHEP} {\bf 12} (2016) 101, [\href{http://arxiv.org/abs/1607.01786}{{\tt
  arXiv:1607.01786}}].

\bibitem{Choi:2016luu}
K.~Choi and S.~H. Im, {\it {Constraints on Relaxion Windows}},  {\em JHEP} {\bf
  12} (2016) 093, [\href{http://arxiv.org/abs/1610.00680}{{\tt
  arXiv:1610.00680}}].

\bibitem{Flacke:2016szy}
T.~Flacke, C.~Frugiuele, E.~Fuchs, R.~S. Gupta, and G.~Perez, {\it
  {Phenomenology of relaxion-Higgs mixing}},
  \href{http://arxiv.org/abs/1610.02025}{{\tt arXiv:1610.02025}}.

\bibitem{McAllister:2016vzi}
L.~McAllister, P.~Schwaller, G.~Servant, J.~Stout, and A.~Westphal, {\it
  {Runaway Relaxion Monodromy}},  \href{http://arxiv.org/abs/1610.05320}{{\tt
  arXiv:1610.05320}}.

\bibitem{Choi:2016kke}
K.~Choi, H.~Kim, and T.~Sekiguchi, {\it {Dynamics of cosmological relaxation
  after reheating}},  \href{http://arxiv.org/abs/1611.08569}{{\tt
  arXiv:1611.08569}}.

\bibitem{Evans:2017bjs}
J.~L. Evans, T.~Gherghetta, N.~Nagata, and M.~Peloso, {\it {Low-Scale D-term
  Inflation and the Relaxion}},  \href{http://arxiv.org/abs/1704.03695}{{\tt
  arXiv:1704.03695}}.

\bibitem{Kusenko:2014lra}
A.~Kusenko, L.~Pearce, and L.~Yang, {\it {Postinflationary Higgs relaxation and
  the origin of matter-antimatter asymmetry}},  {\em Phys. Rev. Lett.} {\bf
  114} (2015), no.~6 061302, [\href{http://arxiv.org/abs/1410.0722}{{\tt
  arXiv:1410.0722}}].

\bibitem{Yang:2015ida}
L.~Yang, L.~Pearce, and A.~Kusenko, {\it {Leptogenesis via Higgs Condensate
  Relaxation}},  {\em Phys. Rev.} {\bf D92} (2015), no.~4 043506,
  [\href{http://arxiv.org/abs/1505.07912}{{\tt arXiv:1505.07912}}].

\bibitem{Kusenko:2014uta}
A.~Kusenko, K.~Schmitz, and T.~T. Yanagida, {\it {Leptogenesis via Axion
  Oscillations after Inflation}},  {\em Phys. Rev. Lett.} {\bf 115} (2015),
  no.~1 011302, [\href{http://arxiv.org/abs/1412.2043}{{\tt arXiv:1412.2043}}].

\bibitem{You:2017kah}
T.~You, {\it {A Dynamical Weak Scale from Inflation}},
  \href{http://arxiv.org/abs/1701.09167}{{\tt arXiv:1701.09167}}.

\bibitem{Arvanitaki:2016xds}
A.~Arvanitaki, S.~Dimopoulos, V.~Gorbenko, J.~Huang, and K.~Van~Tilburg, {\it
  {A small weak scale from a small cosmological constant}},
  \href{http://arxiv.org/abs/1609.06320}{{\tt arXiv:1609.06320}}.

\bibitem{Agugliaro:2016clv}
A.~Agugliaro, O.~Antipin, D.~Becciolini, S.~De~Curtis, and M.~Redi, {\it {UV
  complete composite Higgs models}},  {\em Phys. Rev.} {\bf D95} (2017), no.~3
  035019, [\href{http://arxiv.org/abs/1609.07122}{{\tt arXiv:1609.07122}}].

\bibitem{Joglekar:2012vc}
A.~Joglekar, P.~Schwaller, and C.~E.~M. Wagner, {\it {Dark Matter and Enhanced
  Higgs to Di-photon Rate from Vector-like Leptons}},  {\em JHEP} {\bf 12}
  (2012) 064, [\href{http://arxiv.org/abs/1207.4235}{{\tt arXiv:1207.4235}}].

\bibitem{Manohar:1983md}
A.~Manohar and H.~Georgi, {\it {Chiral Quarks and the Nonrelativistic Quark
  Model}},  {\em Nucl. Phys.} {\bf B234} (1984) 189--212.

\bibitem{German:2001tz}
G.~German, G.~G. Ross, and S.~Sarkar, {\it {Low scale inflation}},  {\em Nucl.
  Phys.} {\bf B608} (2001) 423--450,
  [\href{http://arxiv.org/abs/hep-ph/0103243}{{\tt hep-ph/0103243}}].

\bibitem{Dine:2011ws}
M.~Dine and L.~Pack, {\it {Studies in Small Field Inflation}},  {\em JCAP} {\bf
  1206} (2012) 033, [\href{http://arxiv.org/abs/1109.2079}{{\tt
  arXiv:1109.2079}}].

\bibitem{Iso:2015wsf}
S.~Iso, K.~Kohri, and K.~Shimada, {\it {Dynamical fine-tuning of initial
  conditions for small field inflation}},  {\em Phys. Rev.} {\bf D93} (2016),
  no.~8 084009, [\href{http://arxiv.org/abs/1511.05923}{{\tt
  arXiv:1511.05923}}].

\bibitem{Amaldi:1987fu}
U.~Amaldi, A.~Bohm, L.~S. Durkin, P.~Langacker, A.~K. Mann, W.~J. Marciano,
  A.~Sirlin, and H.~H. Williams, {\it {A Comprehensive Analysis of Data
  Pertaining to the Weak Neutral Current and the Intermediate Vector Boson
  Masses}},  {\em Phys. Rev.} {\bf D36} (1987) 1385.

\bibitem{Costa:1987qp}
G.~Costa, J.~R. Ellis, G.~L. Fogli, D.~V. Nanopoulos, and F.~Zwirner, {\it
  {Neutral Currents Within and Beyond the Standard Model}},  {\em Nucl. Phys.}
  {\bf B297} (1988) 244--286.

\bibitem{Langacker:1991an}
P.~Langacker and M.-x. Luo, {\it {Implications of precision electroweak
  experiments for $M_t$, $\rho_{0}$, $\sin^2\theta_W$ and grand unification}},
  {\em Phys. Rev.} {\bf D44} (1991) 817--822.

\bibitem{Peskin:1991sw}
M.~E. Peskin and T.~Takeuchi, {\it {Estimation of oblique electroweak
  corrections}},  {\em Phys. Rev.} {\bf D46} (1992) 381--409.

\bibitem{Erler:1994fz}
J.~Erler and P.~Langacker, {\it {Implications of high precision experiments and
  the CDF top quark candidates}},  {\em Phys. Rev.} {\bf D52} (1995) 441--450,
  [\href{http://arxiv.org/abs/hep-ph/9411203}{{\tt hep-ph/9411203}}].

\bibitem{Altarelli:1990zd}
G.~Altarelli and R.~Barbieri, {\it {Vacuum polarization effects of new physics
  on electroweak processes}},  {\em Phys. Lett.} {\bf B253} (1991) 161--167.

\bibitem{Altarelli:1991fk}
G.~Altarelli, R.~Barbieri, and S.~Jadach, {\it {Toward a model independent
  analysis of electroweak data}},  {\em Nucl. Phys.} {\bf B369} (1992) 3--32.
  [Erratum: Nucl. Phys.B376,444(1992)].

\bibitem{Grinstein:1991cd}
B.~Grinstein and M.~B. Wise, {\it {Operator analysis for precision electroweak
  physics}},  {\em Phys. Lett.} {\bf B265} (1991) 326--334.

\bibitem{Altarelli:1993sz}
G.~Altarelli, R.~Barbieri, and F.~Caravaglios, {\it {Nonstandard analysis of
  electroweak precision data}},  {\em Nucl. Phys.} {\bf B405} (1993) 3--23.

\bibitem{Barbieri:1999tm}
R.~Barbieri and A.~Strumia, {\it {What is the limit on the Higgs mass?}},  {\em
  Phys. Lett.} {\bf B462} (1999) 144--149,
  [\href{http://arxiv.org/abs/hep-ph/9905281}{{\tt hep-ph/9905281}}].

\bibitem{Barbieri:2004qk}
R.~Barbieri, A.~Pomarol, R.~Rattazzi, and A.~Strumia, {\it {Electroweak
  symmetry breaking after LEP-1 and LEP-2}},  {\em Nucl. Phys.} {\bf B703}
  (2004) 127--146, [\href{http://arxiv.org/abs/hep-ph/0405040}{{\tt
  hep-ph/0405040}}].

\bibitem{ALEPH:2010aa}
{\bf Tevatron Electroweak Working Group, CDF, DELPHI, SLD Electroweak and Heavy
  Flavour Groups, ALEPH, LEP Electroweak Working Group, SLD, OPAL, D0, L3}
  Collaboration, L.~E.~W. Group, {\it {Precision Electroweak Measurements and
  Constraints on the Standard Model}},
  \href{http://arxiv.org/abs/1012.2367}{{\tt arXiv:1012.2367}}.

\bibitem{deBlas:2016ojx}
J.~de~Blas, M.~Ciuchini, E.~Franco, S.~Mishima, M.~Pierini, L.~Reina, and
  L.~Silvestrini, {\it {Electroweak precision observables and Higgs-boson
  signal strengths in the Standard Model and beyond: present and future}},
  {\em JHEP} {\bf 12} (2016) 135, [\href{http://arxiv.org/abs/1608.01509}{{\tt
  arXiv:1608.01509}}].

\bibitem{Maksymyk:1993zm}
I.~Maksymyk, C.~P. Burgess, and D.~London, {\it {Beyond S, T and U}},  {\em
  Phys. Rev.} {\bf D50} (1994) 529--535,
  [\href{http://arxiv.org/abs/hep-ph/9306267}{{\tt hep-ph/9306267}}].

\bibitem{Group:2012gb}
{\bf CDF, D0} Collaboration, T.~E.~W. Group, {\it {2012 Update of the
  Combination of CDF and D0 Results for the Mass of the W Boson}},
  \href{http://arxiv.org/abs/1204.0042}{{\tt arXiv:1204.0042}}.

\bibitem{ALEPH:2005ab}
{\bf SLD Electroweak Group, DELPHI, ALEPH, SLD, SLD Heavy Flavour Group, OPAL,
  LEP Electroweak Working Group, L3} Collaboration, S.~Schael et~al., {\it
  {Precision electroweak measurements on the $Z$ resonance}},  {\em Phys.
  Rept.} {\bf 427} (2006) 257--454,
  [\href{http://arxiv.org/abs/hep-ex/0509008}{{\tt hep-ex/0509008}}].

\bibitem{Burgess:1993mg}
C.~P. Burgess, S.~Godfrey, H.~Konig, D.~London, and I.~Maksymyk, {\it {A Global
  fit to extended oblique parameters}},  {\em Phys. Lett.} {\bf B326} (1994)
  276--281, [\href{http://arxiv.org/abs/hep-ph/9307337}{{\tt hep-ph/9307337}}].

\bibitem{Burgess:1993vc}
C.~P. Burgess, S.~Godfrey, H.~Konig, D.~London, and I.~Maksymyk, {\it {Model
  independent global constraints on new physics}},  {\em Phys. Rev.} {\bf D49}
  (1994) 6115--6147, [\href{http://arxiv.org/abs/hep-ph/9312291}{{\tt
  hep-ph/9312291}}].

\bibitem{Bamert:1994yq}
P.~Bamert and C.~P. Burgess, {\it {Negative S and light new physics}},  {\em Z.
  Phys.} {\bf C66} (1995) 495--502,
  [\href{http://arxiv.org/abs/hep-ph/9407203}{{\tt hep-ph/9407203}}].

\bibitem{Bechtle:2014ewa}
P.~Bechtle, S.~Heinemeyer, O.~St{\aa}l, T.~Stefaniak, and G.~Weiglein, {\it
  {Probing the Standard Model with Higgs signal rates from the Tevatron, the
  LHC and a future ILC}},  {\em JHEP} {\bf 11} (2014) 039,
  [\href{http://arxiv.org/abs/1403.1582}{{\tt arXiv:1403.1582}}].

\bibitem{Eboli:2000ze}
O.~J.~P. Eboli and D.~Zeppenfeld, {\it {Observing an invisible Higgs boson}},
  {\em Phys. Lett.} {\bf B495} (2000) 147--154,
  [\href{http://arxiv.org/abs/hep-ph/0009158}{{\tt hep-ph/0009158}}].

\bibitem{Bechtle:2013xfa}
P.~Bechtle, S.~Heinemeyer, O.~St{\aa}l, T.~Stefaniak, and G.~Weiglein, {\it
  {$HiggsSignals$: Confronting arbitrary Higgs sectors with measurements at the
  Tevatron and the LHC}},  {\em Eur. Phys. J.} {\bf C74} (2014), no.~2 2711,
  [\href{http://arxiv.org/abs/1305.1933}{{\tt arXiv:1305.1933}}].

\bibitem{Bechtle:2008jh}
P.~Bechtle, O.~Brein, S.~Heinemeyer, G.~Weiglein, and K.~E. Williams, {\it
  {HiggsBounds: Confronting Arbitrary Higgs Sectors with Exclusion Bounds from
  LEP and the Tevatron}},  {\em Comput. Phys. Commun.} {\bf 181} (2010)
  138--167, [\href{http://arxiv.org/abs/0811.4169}{{\tt arXiv:0811.4169}}].

\bibitem{Bechtle:2011sb}
P.~Bechtle, O.~Brein, S.~Heinemeyer, G.~Weiglein, and K.~E. Williams, {\it
  {HiggsBounds 2.0.0: Confronting Neutral and Charged Higgs Sector Predictions
  with Exclusion Bounds from LEP and the Tevatron}},  {\em Comput. Phys.
  Commun.} {\bf 182} (2011) 2605--2631,
  [\href{http://arxiv.org/abs/1102.1898}{{\tt arXiv:1102.1898}}].

\bibitem{Bechtle:2013gu}
P.~Bechtle, O.~Brein, S.~Heinemeyer, O.~Stal, T.~Stefaniak, G.~Weiglein, and
  K.~Williams, {\it {Recent Developments in HiggsBounds and a Preview of
  HiggsSignals}},  {\em PoS} {\bf CHARGED2012} (2012) 024,
  [\href{http://arxiv.org/abs/1301.2345}{{\tt arXiv:1301.2345}}].

\bibitem{Bechtle:2013wla}
P.~Bechtle, O.~Brein, S.~Heinemeyer, O.~Stål, T.~Stefaniak, G.~Weiglein, and
  K.~E. Williams, {\it {$\mathsf{HiggsBounds}-4$: Improved Tests of Extended
  Higgs Sectors against Exclusion Bounds from LEP, the Tevatron and the LHC}},
  {\em Eur. Phys. J.} {\bf C74} (2014), no.~3 2693,
  [\href{http://arxiv.org/abs/1311.0055}{{\tt arXiv:1311.0055}}].

\bibitem{MelladoGarcia:2150771}
B.~Mellado~Garcia, P.~Musella, M.~Grazzini, and R.~Harlander, {\it {CERN Report
  4: Part I Standard Model Predictions}},  Tech. Rep.
  LHCHXSWG-DRAFT-INT-2016-008, CERN, May, 2016.

\bibitem{Kawasaki:2004qu}
M.~Kawasaki, K.~Kohri, and T.~Moroi, {\it {Big-Bang nucleosynthesis and
  hadronic decay of long-lived massive particles}},  {\em Phys. Rev.} {\bf D71}
  (2005) 083502, [\href{http://arxiv.org/abs/astro-ph/0408426}{{\tt
  astro-ph/0408426}}].

\bibitem{Jedamzik:2006xz}
K.~Jedamzik, {\it {Big bang nucleosynthesis constraints on hadronically and
  electromagnetically decaying relic neutral particles}},  {\em Phys. Rev.}
  {\bf D74} (2006) 103509, [\href{http://arxiv.org/abs/hep-ph/0604251}{{\tt
  hep-ph/0604251}}].

\bibitem{Jedamzik:2009uy}
K.~Jedamzik and M.~Pospelov, {\it {Big Bang Nucleosynthesis and Particle Dark
  Matter}},  {\em New J. Phys.} {\bf 11} (2009) 105028,
  [\href{http://arxiv.org/abs/0906.2087}{{\tt arXiv:0906.2087}}].

\bibitem{Cyburt:2002uv}
R.~H. Cyburt, J.~R. Ellis, B.~D. Fields, and K.~A. Olive, {\it {Updated
  nucleosynthesis constraints on unstable relic particles}},  {\em Phys. Rev.}
  {\bf D67} (2003) 103521, [\href{http://arxiv.org/abs/astro-ph/0211258}{{\tt
  astro-ph/0211258}}].

\bibitem{Schwaller:2015gea}
P.~Schwaller, D.~Stolarski, and A.~Weiler, {\it {Emerging Jets}},  {\em JHEP}
  {\bf 05} (2015) 059, [\href{http://arxiv.org/abs/1502.05409}{{\tt
  arXiv:1502.05409}}].

\bibitem{Cohen:2015toa}
T.~Cohen, M.~Lisanti, and H.~K. Lou, {\it {Semivisible Jets: Dark Matter
  Undercover at the LHC}},  {\em Phys. Rev. Lett.} {\bf 115} (2015), no.~17
  171804, [\href{http://arxiv.org/abs/1503.00009}{{\tt arXiv:1503.00009}}].

\bibitem{Carloni:2011kk}
L.~Carloni, J.~Rathsman, and T.~Sjostrand, {\it {Discerning Secluded Sector
  gauge structures}},  {\em JHEP} {\bf 04} (2011) 091,
  [\href{http://arxiv.org/abs/1102.3795}{{\tt arXiv:1102.3795}}].

\bibitem{Witten:1979kh}
E.~Witten, {\it {Baryons in the 1/n Expansion}},  {\em Nucl. Phys.} {\bf B160}
  (1979) 57--115.

\bibitem{Aaboud:2016yus}
{\bf ATLAS} Collaboration, M.~Aaboud et~al., {\it {Measurement of the
  $W^{\pm}Z$ boson pair-production cross section in $pp$ collisions at
  $\sqrt{s}=13$ TeV with the ATLAS Detector}},  {\em Phys. Lett.} {\bf B762}
  (2016) 1--22, [\href{http://arxiv.org/abs/1606.04017}{{\tt
  arXiv:1606.04017}}].

\bibitem{Khachatryan:2016tgp}
{\bf CMS} Collaboration, V.~Khachatryan et~al., {\it {Measurement of the WZ
  production cross section in pp collisions at $\sqrt(s) =$ 13 TeV}},  {\em
  Phys. Lett.} {\bf B766} (2017) 268--290,
  [\href{http://arxiv.org/abs/1607.06943}{{\tt arXiv:1607.06943}}].

\bibitem{Aad:2016ett}
{\bf ATLAS} Collaboration, G.~Aad et~al., {\it {Measurements of $W^\pm Z$
  production cross sections in $pp$ collisions at $\sqrt{s} = 8$ TeV with the
  ATLAS detector and limits on anomalous gauge boson self-couplings}},  {\em
  Phys. Rev.} {\bf D93} (2016), no.~9 092004,
  [\href{http://arxiv.org/abs/1603.02151}{{\tt arXiv:1603.02151}}].

\bibitem{Khachatryan:2016poo}
{\bf CMS} Collaboration, V.~Khachatryan et~al., {\it {Measurement of the WZ
  production cross section in pp collisions at $\sqrt{s}$ = 7 and 8 TeV and
  search for anomalous triple gauge couplings at $\sqrt{s}$ = 8 TeV}},  {\em
  Submitted to: Eur. Phys. J. C} (2016)
  [\href{http://arxiv.org/abs/1609.05721}{{\tt arXiv:1609.05721}}].

\bibitem{Alloul:2013bka}
A.~Alloul, N.~D. Christensen, C.~Degrande, C.~Duhr, and B.~Fuks, {\it
  {FeynRules 2.0 - A complete toolbox for tree-level phenomenology}},  {\em
  Comput. Phys. Commun.} {\bf 185} (2014) 2250--2300,
  [\href{http://arxiv.org/abs/1310.1921}{{\tt arXiv:1310.1921}}].

\bibitem{Alwall:2014hca}
J.~Alwall, R.~Frederix, S.~Frixione, V.~Hirschi, F.~Maltoni, O.~Mattelaer,
  H.~S. Shao, T.~Stelzer, P.~Torrielli, and M.~Zaro, {\it {The automated
  computation of tree-level and next-to-leading order differential cross
  sections, and their matching to parton shower simulations}},  {\em JHEP} {\bf
  07} (2014) 079, [\href{http://arxiv.org/abs/1405.0301}{{\tt
  arXiv:1405.0301}}].

\bibitem{Sjostrand:2006za}
T.~Sjostrand, S.~Mrenna, and P.~Z. Skands, {\it {PYTHIA 6.4 Physics and
  Manual}},  {\em JHEP} {\bf 05} (2006) 026,
  [\href{http://arxiv.org/abs/hep-ph/0603175}{{\tt hep-ph/0603175}}].

\bibitem{deFavereau:2013fsa}
{\bf DELPHES 3} Collaboration, J.~de~Favereau, C.~Delaere, P.~Demin,
  A.~Giammanco, V.~Lemaître, A.~Mertens, and M.~Selvaggi, {\it {DELPHES 3, A
  modular framework for fast simulation of a generic collider experiment}},
  {\em JHEP} {\bf 02} (2014) 057, [\href{http://arxiv.org/abs/1307.6346}{{\tt
  arXiv:1307.6346}}].

\bibitem{Cacciari:2011ma}
M.~Cacciari, G.~P. Salam, and G.~Soyez, {\it {FastJet User Manual}},  {\em Eur.
  Phys. J.} {\bf C72} (2012) 1896, [\href{http://arxiv.org/abs/1111.6097}{{\tt
  arXiv:1111.6097}}].

\bibitem{ATLAS:2016iqc}
{\bf ATLAS} Collaboration, T.~A. collaboration, {\it {Electron efficiency
  measurements with the ATLAS detector using the 2015 LHC proton-proton
  collision data}}, .

\bibitem{CMS:eff}
```{C}uts in {C}ategories" ({C}i{C}) {E}lectron {I}dentification.''
  \url{https://twiki.cern.ch/twiki/bin/view/CMSPublic/SWGuideCategoryBasedElectronID}.
\newblock Accessed: 2017-03-15.

\bibitem{Aad:2016jkr}
{\bf ATLAS} Collaboration, G.~Aad et~al., {\it {Muon reconstruction performance
  of the ATLAS detector in proton–proton collision data at $\sqrt{s}$ =13
  TeV}},  {\em Eur. Phys. J.} {\bf C76} (2016), no.~5 292,
  [\href{http://arxiv.org/abs/1603.05598}{{\tt arXiv:1603.05598}}].

\bibitem{Beenakker:1996ed}
W.~Beenakker, R.~Hopker, and M.~Spira, {\it {PROSPINO: A Program for the
  production of supersymmetric particles in next-to-leading order QCD}},
  \href{http://arxiv.org/abs/hep-ph/9611232}{{\tt hep-ph/9611232}}.

\bibitem{Read:2002hq}
A.~L. Read, {\it {Presentation of search results: The CL(s) technique}},  {\em
  J. Phys.} {\bf G28} (2002) 2693--2704. [,11(2002)].

\bibitem{Junk:1999kv}
T.~Junk, {\it {Confidence level computation for combining searches with small
  statistics}},  {\em Nucl. Instrum. Meth.} {\bf A434} (1999) 435--443,
  [\href{http://arxiv.org/abs/hep-ex/9902006}{{\tt hep-ex/9902006}}].

\bibitem{CMS:2017fdz}
{\bf CMS} Collaboration, C.~Collaboration, {\it {Search for electroweak
  production of charginos and neutralinos in multilepton final states in pp
  collision data at $\sqrt{s}=13~\mathrm{TeV}$}}, .

\bibitem{Staub:2013tta}
F.~Staub, {\it {SARAH 4 : A tool for (not only SUSY) model builders}},  {\em
  Comput. Phys. Commun.} {\bf 185} (2014) 1773--1790,
  [\href{http://arxiv.org/abs/1309.7223}{{\tt arXiv:1309.7223}}].

\bibitem{Degrassi:2012ry}
G.~Degrassi, S.~Di~Vita, J.~Elias-Miro, J.~R. Espinosa, G.~F. Giudice,
  G.~Isidori, and A.~Strumia, {\it {Higgs mass and vacuum stability in the
  Standard Model at NNLO}},  {\em JHEP} {\bf 08} (2012) 098,
  [\href{http://arxiv.org/abs/1205.6497}{{\tt arXiv:1205.6497}}].

\bibitem{Isidori:2001bm}
G.~Isidori, G.~Ridolfi, and A.~Strumia, {\it {On the metastability of the
  standard model vacuum}},  {\em Nucl. Phys.} {\bf B609} (2001) 387--409,
  [\href{http://arxiv.org/abs/hep-ph/0104016}{{\tt hep-ph/0104016}}].

\bibitem{Bandyopadhyay:2016oif}
P.~Bandyopadhyay and R.~Mandal, {\it {Vacuum stability in an extended standard
  model with a leptoquark}},  {\em Phys. Rev.} {\bf D95} (2017), no.~3 035007,
  [\href{http://arxiv.org/abs/1609.03561}{{\tt arXiv:1609.03561}}].

\bibitem{LEP:ch1}
``{LEP2 SUSY Working Group}.'' ALEPH, DELPHI, L3 and OPAL experiments, note
  LEPSUSYWG/01-03.1.
\newblock \url{http://lepsusy.web.cern.ch/lepsusy}.

\bibitem{LEP:ch2}
``{LEP2 SUSY Working Group}.'' ALEPH, DELPHI, L3 and OPAL experiments, note
  LEPSUSYWG/02-04.1.
\newblock \url{http://lepsusy.web.cern.ch/lepsusy}.

\end{thebibliography}\endgroup
\bibliographystyle{JHEP}

\end{document}